\documentclass[preprint,12pt]{elsarticle}

\usepackage{amssymb}
\usepackage{amsmath}
\usepackage[bookmarks=true]{hyperref}
\usepackage{etex} % solve "No room for a new dimen"
\usepackage{caption}
\usepackage{subcaption}

\usepackage{ntheorem}
\theoremstyle{break}

\usepackage{graphicx}
\usepackage{verbatim}
\usepackage{tikz}
\usetikzlibrary{arrows,positioning,shadows,matrix,calc,shapes}  
\usepackage{pgf,pgfarrows,pgfnodes,pgfautomata,pgfheaps,pgfplots}
\usepackage{pst-plot}   %coordinate system

\usepackage[ruled,vlined,linesnumbered]{algorithm2e}

\usepackage{booktabs} % For professional table styling
\usepackage{array} % For column formatting
\newcolumntype{L}{>{\raggedright\arraybackslash}p{0.2\textwidth}} % Left-aligned column
\newcolumntype{R}{>{\raggedright\arraybackslash}p{0.6\textwidth}} % Description column

\usepackage{wrapfig}

\usepackage{url}
\urldef{\mailsa}\path|{alfred.hofmann, ursula.barth, ingrid.haas, frank.holzwarth,|
	\urldef{\mailsb}\path|anna.kramer, leonie.kunz, christine.reiss, nicole.sator,|
	\urldef{\mailsc}\path|erika.siebert-cole, peter.strasser, lncs}@springer.com|

\newdefinition{definition}{Definition}
\newdefinition{example}{Example}
\newtheorem{theorem}{Theorem}
\newtheorem{proposition}{Proposition}
\newtheorem{lemma}{Lemma}
\newtheorem{corollary}{Corollary}
\newtheorem{assumption}{Assumption}
\newtheorem{remark}{Remark}
\newproof{proof}{Proof}

\journal{Information Sciences}

\begin{document}
	
	%% ========== Journal Pre-proof Cover Page ==========
	\begin{titlepage}
		\centering
		\vspace*{2cm}
		
		{\LARGE Journal Pre-proof\par}
		\vspace{1.5cm}
		
		{\Huge\bfseries Achieving Dalenius' Goal of Data Privacy with Practical Assumptions\par}
		\vspace{1.5cm}
		
		{\Large Genqiang Wu, Xianyao Xia, Yeping He\par}
		\vspace{2cm}
		
		\begin{minipage}{\textwidth}
			\centering
			PII: S0020-0255(25)01172-7\\
			DOI: \url{https://doi.org/10.1016/j.ins.2025.123035}\\
			Reference: INS 123035\\
			\vspace{0.5cm}
			To appear in: \textit{Information Sciences}
		\end{minipage}
		
		\vfill
		
		\begin{minipage}{\textwidth}
			\centering
			Received Date: 5 May 2025\\
			Revised Date: 23 December 2025\\
			Accepted Date: 23 December 2025
		\end{minipage}
		
		\vfill
		
		\begin{minipage}{0.9\textwidth}
			\footnotesize
			\raggedright
			\textit{This is a PDF of an article that has undergone enhancements after acceptance, such as the addition of a cover page and metadata, and formatting for readability. This version will undergo additional copyediting, typesetting and review before it is published in its final form. As such, this version is no longer the Accepted Manuscript, but it is not yet the definitive Version of Record; we are providing this early version to give early visibility of the article. Please note that Elsevier's sharing policy for the Published Journal Article applies to this version, see: \url{https://www.elsevier.com/about/policies-and-standards/sharing\#4-published-journal-article}. Please also note that, during the production process, errors may be discovered which could affect the content, and all legal disclaimers that apply to the journal pertain.}
		\end{minipage}
		
		\vfill
		
		\textcopyright\ 2025 Elsevier Inc. All rights are reserved, including those for text and data mining, AI training, and similar technologies.
		
		\vspace*{1cm}
	\end{titlepage}
	%% ========== End of Cover Page ==========
	
	\newpage
	
	\begin{frontmatter}
		
		
		
		\title{Achieving Dalenius' Goal of Data Privacy with Practical Assumptions}
		
		
		\author[1,2]{Genqiang Wu\fnref{fn1}\corref{cor1}}
		\ead{wahaha2000@icloud.com}
		\fntext[fn1]{This work was partially done when the first author was with Chongqing University of Technology and with Institute of Software Chinese Academy of Sciences.}
		
		\author[3]{Xianyao Xia}
		\ead{xianyao@hotmail.com}
		
		\author[3]{Yeping He}
		\ead{yeping@nfs.iscas.ac.cn}
		
		\affiliation[1]{organization={School of Information Engineering and Artificial Intelligence, Lanzhou University of Finance and Economics}, 
			addressline={No. 4 Weile Avenue, Heping Campus},
			city={Lanzhou},
			postcode={730101},
			%state={Gansu}, 
			country={China}}

		\affiliation[2]{organization={Gansu Province Key Laboratory of Smart Commerce},
			%addressline={},
			city={Lanzhou},
			%postcode={},
			%state={},
			country={China}}
		
		\affiliation[3]{organization={Institute of Software, Chinese Academy of Sciences},
			%addressline={},
			city={Beijing},
			%postcode={},
			%state={},
			country={China}}
		\cortext[cor1]{Corresponding author}  % 标记显示为脚注

		%% Abstract
		\begin{abstract}
			Current differential privacy frameworks face significant challenges: vulnerability to correlated data attacks and suboptimal utility-privacy tradeoffs. To address these limitations, we establish a novel information-theoretic foundation for Dalenius' privacy vision using Shannon's perfect secrecy framework. By leveraging the fundamental distinction between cryptographic systems (small secret keys) and privacy mechanisms (massive datasets), we replace differential privacy's restrictive independence assumption with practical partial knowledge constraints ($H(X) \geq b$). 
			
			We propose an information privacy framework achieving Dalenius security with quantifiable utility-privacy tradeoffs. Crucially, we prove that foundational mechanisms---random response, exponential, and Gaussian channels---satisfy Dalenius' requirements while preserving group privacy and composition properties. Our channel capacity analysis reduces infinite-dimensional evaluations to finite convex optimizations, enabling direct application of information-theoretic tools. 
			
			Empirical evaluation demonstrates that individual channel capacity (maximal information leakage of each individual) decreases with increasing entropy constraint $b$, and our framework achieves superior utility-privacy tradeoffs compared to classical differential privacy mechanisms under equivalent privacy guarantees. The framework is extended to computationally bounded adversaries via Yao's theory, unifying cryptographic and statistical privacy paradigms. Collectively, these contributions provide a theoretically grounded path toward practical, composable privacy---subject to future resolution of the tradeoff characterization---with enhanced resilience to correlation attacks.
		\end{abstract}

%% Graphical abstract
\begin{graphicalabstract}	
	\tikzstyle{DataPrivacy} = [rectangle, rounded corners, minimum width=3cm, minimum height=1cm,text centered, draw=black, fill=red!10!white]
	\tikzstyle{Cryptography} = [rectangle, rounded corners, minimum width=3cm, minimum height=1cm,text centered, draw=black, fill=green!10!white]
	\tikzstyle{arrow} = [thick,->,>=stealth]
	\tikzstyle{bluearrow} = [thick,->,>=stealth, draw=blue]
	\tikzstyle{greenarrow} = [thick,->,>=stealth, draw=green]
	\begin{figure}[!b]
		\begin{tikzpicture}[->, %thick, 
			node distance=2.5cm,
			%every node/.style={scale=0.8, draw=none}, 
			scale=0.6, transform shape]
			% Define nodes
			\node (1a) [DataPrivacy, align=center] at (0,0) {Data Privacy \\ dataset $X := (X_1,\ldots, X_n)$};
			\node (1b) [Cryptography, align=center] at (8,0) {Cryptography \\ secret key $X$};
			\node (2a) [DataPrivacy, align=center] at (0,-3) {Dalenius Security \\ $I(X_i;Y)\le \epsilon$ \\ arbitrary adversaries};
			\node (2b) [Cryptography, align=center] at (8,-3) {Perfect Secrecy \\ $I(X;Y) = 0 $ \\ arbitrary adversaries};
			\node (3a) [DataPrivacy, align=center] at (-8,-6) {Differential Privacy \\ $I_{\infty}(X_i;Y)\le\epsilon$ \\ independence knowledge adversaries: \\ $X_1,\ldots,X_n$ independent};
			\node (3b) [Cryptography, align=center] at (8,-6) {Semantic/Computational Security\\ $I_c(X;Y) = 0 $ \\ computational-bounded adversaries};
			\node (4a) [DataPrivacy, align=center] at (0,-11) {Information Privacy\\ $I(X_i;Y)\le\epsilon$ \\ partial knowledge adversaries: \\ $H(X) \ge b \;\;\; (b>0)$};
			\draw [arrow] (1a) -- (2a);
			\draw [arrow] (1b) -- (2b);
			\draw [arrow] (2a) -- (3a);
			\draw [arrow] (2b) -- (3b);
			\draw [arrow] (3a) -- (4a);
			\draw [bluearrow] (3b) -- node[pos=0.5, sloped, below, align=center, fill=gray!30] {small secret \\ key $X$} (4a);
			\draw [bluearrow] (3a) -- node[pos=0.5, sloped, below, align=center, fill=gray!30] {massive \\ dataset  $X$} (4a);
			\draw [greenarrow] (2b) -- (2a);
			\draw [greenarrow] (2a) -- (4a);
		\end{tikzpicture}
		\caption{
			A conceptual mapping between cryptographic security and data privacy frameworks. 
			Left branch illustrates the evolution of privacy models: 
			(1) From Dalenius security extending perfect secrecy, 
			(2) Through differential privacy under independence assumptions, 
			(3) To our proposed information privacy framework with partial knowledge constraints ($H(X) \geq b$). 
			Right branch depicts cryptographic counterparts: 
			(1) Perfect secrecy via information-theoretic approaches, 
			(2) Semantic security under computational constraints. 
			The convergence diamond highlights our core innovation: 
			Blue arrows with gray labels highlight key distinctions between small-key cryptography and big-data privacy mechanisms that motivate partial knowledge constraints. 
			Green arrows indicate information-theoretic consistency between Shannon's secrecy, Dalenius security, and information privacy.
		}
	\end{figure}
\end{graphicalabstract}


\begin{highlights}
	\item Shannon-based Dalenius' privacy framework with partial knowledge constraint
	\item Replaced DP's independence assumption with entropy-based adversary constraint 
	\item Composable foundational mechanisms (RR, Exp, Gauss) satisfy Dalenius security
	\item Reduced channel capacity analysis to finite convex optimization
	\item Unified information-theoretic \& computational privacy via Yao's theory
\end{highlights}

%% Keywords

\end{frontmatter}


\section{Introduction} \label{section-introduction}

The exponential growth of data collection, particularly personal information, has been significantly accelerated by advancements in artificial intelligence (AI) ~\cite{2023arXiv230602781G} and big data analytics~\cite{mayer2013big}. While these technologies drive innovation through data-driven decision making, they simultaneously introduce substantial privacy risks. Sensitive personal attributes---including location data, health records, and financial information---remain vulnerable to re-identification attacks and security breaches, potentially exposing individuals to surveillance, discrimination, and identity theft~\cite{DBLP:conf/sp/NarayananS08}. Notably, recent studies reveal vulnerabilities even in modern generative models like GPT-2 (Generative Pre-trained Transformer 2), which are susceptible to training data extraction attacks~\cite{carlini2021extracting}. These enable adversaries to recover verbatim text sequences containing sensitive information through black-box queries~\cite{carlini2021extracting} or unprompted outputs~\cite{lee2022deduplicating}. Furthermore, membership inference attacks can determine whether specific instances were used in model training~\cite{hu2022membership}.

Current regulatory frameworks such as the GDPR (General Data Protection Regulation) \cite{voigt2017eu} and the EU AI Act \cite{europeanunion2024artificial} attempt to address these challenges through legislative measures. However, technical privacy protection mechanisms remain essential to complement these regulations. The field of data privacy protection \cite{near2025guidelines} focuses on deriving statistically valid insights while preserving individual privacy in datasets containing sensitive information. This challenge has become increasingly critical in our data-driven era.

Among existing approaches, differential privacy (DP) \cite{DBLP:journals/fttcs/DworkR14} has emerged as the dominant paradigm due to two key characteristics: (1) its semantic-agnostic privacy definition, analogous to Shannon's information theory \cite{Shannon1948}, which decouples privacy guarantees from specific data semantics; and (2) its composability properties, enabling modular construction of complex private algorithms from simpler components \cite{DBLP:journals/fttcs/DworkR14}. These features have driven widespread adoption across academia \cite{zhao2024scenario} and industry \cite{DBLP:conf/ccs/ErlingssonPK14}.

Despite its success, differential privacy faces two fundamental limitations. First, it demonstrates vulnerabilities to correlation-aware adversarial attacks \cite{DBLP:journals/concurrency/0055ZL022}, particularly when adversaries possess knowledge of inter-record correlations – a common scenario in real-world datasets. Second, its utility-privacy tradeoff is often suboptimal in many applications \cite{DBLP:conf/crypto/GehrkeHLP12}, failing to leverage the ``crowd blending'' phenomenon where privacy guarantees naturally strengthen with larger datasets \cite{DBLP:journals/ijufks/Sweene02}. These limitations have spurred development of alternative models including Pufferfish \cite{DBLP:journals/tods/KiferM14}, coupled-worlds privacy \cite{DBLP:conf/focs/BassilyGKS13}, zero-knowledge privacy \cite{DBLP:conf/tcc/GehrkeLP11}, concentrated differential privacy (CDP) \cite{DBLP:journals/corr/DworkR16}, and R\'{e}nyi differential privacy (RDP) \cite{DBLP:conf/tcc/BunS16} among others, though none has achieved both superior security and utility while maintaining composability \cite{DBLP:journals/popets/DesfontainesP20}.

This work revisits the foundational question: Can we achieve Dalenius' original privacy vision~\cite{dalenius1977towards} – that no individual's information should be learnable from a dataset beyond what's possible without accessing it – while maintaining practical utility \cite{DBLP:conf/icalp/Dwork06}? To answer this question, we construct a new data privacy framework by Shannon's information-theoretic framework to cryptography \cite{6769090}. This framework is powerful enough to model many popular data privacy definitions, including differential privacy, Dalenius security, Pufferfish privacy etc. By applying this framework, we make the following contributions, which significantly advance beyond existing privacy frameworks:

\textbf{First}, unlike prior work that either assumes independence (differential privacy) or requires complex correlation specifications (Pufferfish), we establish formal equivalence between differential privacy and Dalenius' goal under the independence assumption, then introduce a novel \textit{partial knowledge} assumption ($H(X) \geq b$) that is both more practical than independence and more tractable than full correlation modeling. This represents a fundamental shift from modeling data correlations to constraining adversary knowledge.

\textbf{Second}, we overcome a key computational barrier in privacy analysis by developing a complexity reduction technique that transforms infinite-dimensional individual channel capacity evaluations into finite convex optimizations. This technical innovation enables practical application of information-theoretic tools where previous approaches faced computational intractability.

\textbf{Third}, we establish fundamental privacy properties including post-processing invariance, group privacy preservation, and composition properties that are provably maintained under our framework. Crucially, we demonstrate that these properties hold without relying on the independence assumptions required by differential privacy or the complex specifications of Pufferfish.

\textbf{Fourth}, our extension to computational settings using Yao's theory provides the first unified framework that seamlessly bridges information-theoretic and computational privacy paradigms. This unification naturally incorporates both cryptographic primitives and computational differential privacy as special cases, offering a more comprehensive foundation than existing compartmentalized approaches.

\tikzstyle{DataPrivacy} = [rectangle, rounded corners, minimum width=3cm, minimum height=1cm,text centered, draw=black, fill=red!10!white]
\tikzstyle{Cryptography} = [rectangle, rounded corners, minimum width=3cm, minimum height=1cm,text centered, draw=black, fill=green!10!white]
\tikzstyle{arrow} = [thick,->,>=stealth]
\tikzstyle{bluearrow} = [thick,->,>=stealth, draw=blue]
\tikzstyle{greenarrow} = [thick,->,>=stealth, draw=green]
\begin{figure}[!b]
	\begin{tikzpicture}[->, %thick, 
		node distance=2.5cm,
		%every node/.style={scale=0.8, draw=none}, 
		scale=0.6, transform shape]
		% Define nodes
		\node (0C) [DataPrivacy, align=center] at (-4,3) {Section 2 \\ IP Model};
		\node (1C) [DataPrivacy, align=center] at (0,0) {Section 2.2 \\ IP with $H(X) \ge b$};
		\node (1L) [DataPrivacy, align=center] at (-8,0) {Section 2.1 \\ DP/IP with idependence assumption};
		\node (2C) [DataPrivacy, align=center] at (0,-3) {Section 3 \\ evaluating individual \\ channel capacity};
		\node (2L) [DataPrivacy, align=center] at (-8,-3) {Section 5 \\ properties of IP};
		\node (2R) [DataPrivacy, align=center] at (6,-3) {Section 6 \\ computational IP};
		\node (3C) [DataPrivacy, align=center] at (0,-6) {Section 4 \\ fundamental privacy channels};
		\node (3L1) [DataPrivacy, align=center] at (-11,-6) {Section 5.1 \\ group privacy property};
		\node (3L2) [DataPrivacy, align=center] at (-6,-6) {Section 5.2 \\ composition property};
		\node (3R1) [DataPrivacy, align=center] at (6,-6) {Section 7 \\ evaluation};		
		\node (4C1) [DataPrivacy, align=center] at (-9,-9) {Section 4.1 \\ binary-symmetric privacy channel};		
		\node (4C2) [DataPrivacy, align=center] at (-1,-9) {Section 4.3 \\ exponential privacy channel};		
		\node (4C3) [DataPrivacy, align=center] at (6,-9) {Section 4.4 \\ Gaussian privacy channel};
		\draw [arrow] (0C) -- (1L);
		\draw [arrow] (0C) -- (1C);
		\draw [arrow] (1C) -- (2L);
		\draw [arrow] (1C) -- (2C);
		\draw [arrow] (1C) -- (2R);
		\draw [arrow] (2L) -- (3L1);
		\draw [arrow] (2L) -- (3L2);
		\draw [arrow] (2C) -- (3C);	
		\draw [arrow] (2C) -- (3R1);	
		\draw [arrow] (3C) -- (4C1);		
		\draw [arrow] (3C) -- (4C2);		
		\draw [arrow] (3C) -- (4C3);		
		% \draw [bluearrow] (3a) -- node[pos=0.5, sloped, below, align=center, fill=gray!30] {massive \\ dataset  $X$} (4a);
		% \draw [greenarrow] (2a) -- (4a);
	\end{tikzpicture}
	\caption{The flowchart of this paper's structure. Section \ref{section:prelim} formalizes our information-theoretic framework and assumptions. Subsequent sections detail our core contributions in channel capacity analysis (Sections \ref{sec-individual-channel-capacity}-\ref{sec:some-channels}), group privacy and composition properties (Section \ref{sec-group-composition-privacy}), computational extensions (Section \ref{sec-computational-variants}), and experimental evaluations (Section \ref{sec:evaluation}). We provide a comprehensive analysis of related works in Section \ref{section:related-works} and finally summarize the key findings in Section \ref{section:conclusion}. }
	\label{fig:flowchart_architecture}
\end{figure}

\textbf{Fifth}, we present comprehensive empirical evaluation demonstrating that our framework achieves superior utility-privacy tradeoffs compared to classical differential privacy mechanisms under equivalent privacy guarantees, while providing enhanced resilience to correlation attacks that plague traditional differential privacy.

\emph{It is important to note that the algorithms for evaluating quantities like individual channel capacity under entropy constraints involve complex, high-dimensional non-convex optimization. Their detailed description and analysis ---spanning over 40 pages alone---are highly technical and would exceed the scope and length constraints of this paper. We thus focus here on the core privacy model construction and present experimental validation of key relationships, while the algorithmic developments and associated theoretical guarantees will be presented in a separate, forthcoming paper.}

Our results collectively demonstrate a privacy framework achieving Dalenius' vision under practical knowledge assumptions, with composable guarantees and quantifiable utility-privacy tradeoffs. 

Figure \ref{fig:flowchart_architecture} illustrates the flowchart of this paper's structure.

\begin{table}[!tbh]
	\caption{Nomenclature}
	\label{tab:nomenclature}
	\centering
	\footnotesize
	\begin{tabular}{@{} p{0.26\textwidth} p{0.49\textwidth} p{0.19\textwidth} @{}}
		\toprule
		\textbf{Symbol/Term} & \textbf{Description} & \textbf{Location} \\
		\midrule
		\multicolumn{3}{@{}l}{\textit{Dataset $\&$ Adversary}} \\
		$n$ & Number of records in the dataset & Sec \ref{section:prelim}\\
		$x_i \in \mathcal{X}_i$ & Private data of individual $i$ & Sec \ref{section:prelim} \\
		${X =}  (X_1, \dots, X_n)$ & Adversary or its knowledge to dataset & Sec \ref{section:prelim} \\
		$\mathcal{X} = \prod_{i=1}^n \mathcal{X}_i$ & Universe of possible datasets & Sec \ref{section:prelim} \\
		$X_{-i}$ & $(X_1,\ldots,X_{i-1}, X_{i+1}, \ldots, X_n)$ & Sec \ref{section:prelim} \\
		$x_{-i}$ & $(x_1,\ldots,x_{i-1}, x_{i+1}, \ldots, x_n)$ & Sec \ref{section:prelim} \\
		\midrule
		\multicolumn{3}{@{}l}{\textit{Privacy Mechanisms}} \\
		$f: \mathcal{X} \rightarrow \mathcal{Y}$ & Deterministic query function & Def \ref{def:privacy_mechanism} (Sec \ref{section:prelim})\\
		$p(y \mid x)$ & Privacy mechanism/channel & Def \ref{def:privacy_mechanism}  (Sec \ref{section:prelim})\\
		$\epsilon$ & Privacy budget & Def \ref{def:dalenius_security} (Sec \ref{section:prelim})\\
		$\Delta(\mathcal{X})$ & Probability simplex over $\mathcal{X}$ & Def \ref{def:probability_simplex} (Sec \ref{section:prelim})\\
		$X \in \Delta(\mathcal{X})$ & the probability of $X$ is in $\Delta(\mathcal{X})$ & Sec \ref{section:prelim}\\
		$\mathbb P $ & An adversary class, $\mathbb P\subseteq \Delta(\mathcal{X})$ & Def \ref{def:individual_channel_capacity} (Sec \ref{section:prelim})\\
		$\mathbb P_{\text{ind}} $ & \scriptsize{Adversary class with product distributions} & Def \ref{def:max_information_privacy} (Sec \ref{section:prelim})\\
		$\Delta(\mathcal X)_b$ & Adversary class with $H(X) \ge b$ & Assumption \ref{assum:bounded_knowledge}  \\
		\midrule
		\multicolumn{3}{@{}l}{\textit{Information Theory}} \\
		$I(X_i; Y)$ & Shannon mutual information & Def \ref{def:dalenius_security} (Sec \ref{section:prelim})\\
		$I_\infty(X_i; Y)$ & Max-mutual information & Def \ref{def:max-MI} (Sec \ref{section:prelim})\\
		$H(X)$ & Shannon entropy of $X$ & Assumption \ref{assum:bounded_knowledge} \\
		$C_1^{\mathbb P}$ & Individual channel capacity w.r.t. $\mathbb P$ & Def \ref{def:individual_channel_capacity} (Sec \ref{section:prelim}) \\
		$C_{1,\infty}^{\mathbb P}$ & Max-individual channel capacity & Def \ref{def:max_information_privacy} (Sec \ref{subsec:dp-model}) \\
		$C_1^b, C_1$ & $C_1^b = \max_{i \in [n], p(x) \in \Delta(\mathcal X)_b} I(X_i;Y) $, $C_1=C_1^0$ & Def \ref{definition-2} (Sec \ref{subsec:new-model}) \\
		$\delta = g(b)$ & Utility-privacy balance function & Def \ref{definition-2} (Sec \ref{subsec:new-model}) \\
		$\mathbb Q_i$   & the set of all transition matrix $p(y|x_i)$ &  Sec \ref{subsec-simple-example} \\
		$\mathbb S_i $   & subset of $\mathbb Q_i$ with degenerate $p(x_{-i}|x_i)$ &  Sec \ref{subsec-simple-example} \\
		\midrule
		\multicolumn{3}{@{}l}{\textit{Privacy Models}} \\
		$\epsilon$-\footnotesize{Dalenius secure} & $I(X_i; Y) \leq \epsilon$ for all practical $X$ & Def \ref{def:dalenius_security} (Sec 2) \\
		$\epsilon$-\scriptsize{Information privacy} & $C_1^{\mathbb P} \leq \epsilon$ w.r.t. $\mathbb P$ & Def \ref{def:information_privacy} (Sec 2) \\
		$\epsilon$\scriptsize{-Max-information privacy} & $C_{1,\infty}^{\mathbb P_{\text{ind}}} \leq \epsilon$ w.r.t $\mathbb P_{\text{ind}}$ & Def \ref{def:max_information_privacy} (Sec 2.1) \\
		\midrule
		\multicolumn{3}{@{}l}{\textit{Acronyms}} \\
		DP & Differential Privacy & Sec \ref{section:prelim} \\
		IP & Information Privacy & Sec \ref{section:prelim} \\
		CDP & Concentrated Differential Privacy & Sec \ref{section:related-works} \\
		RDP & Renyi Differential Privacy & Sec \ref{section:related-works} \\
		\bottomrule
	\end{tabular}
\end{table}

\section{The Information Privacy Model}\label{section:prelim}

This section establishes the foundational framework for information privacy (IP) modeling. We begin by formalizing Dalenius' desideratum using Shannon's perfect secrecy criterion, extending Shannon's information-theoretic model of encryption systems \cite{6769090}. Subsequent sections will analyze the differential privacy model's security limitations under independence assumptions (Section~\ref{subsec:dp-model}), and propose a novel privacy model with relaxed assumptions for enhanced security and utility flexibility (Section~\ref{subsec:new-model}).

\paragraph{Notational Convention} Our notation follows standard information theory conventions from \cite{DBLP:books/daglib/0016881}. Consider a dataset containing $n\geq 1$ records where each record $x_i \in \mathcal{X}_i$ corresponds to an individual's private data. Let $X = (X_1,\ldots,X_n)$ represent an adversary's probabilistic uncertainty about the dataset, with $\mathcal{X} = \prod_{i=1}^n \mathcal{X}_i$ denoting the complete record universe. We define marginal components as $\mathcal{X}_{-i} = \prod_{j\neq i} \mathcal{X}_j$ and corresponding random variable tuples $X_{-i} = (X_1,\ldots,X_{i-1},X_{i+1},\ldots,X_n)$. Table \ref {tab:nomenclature} presents the nomenclature of this paper.

\begin{definition}[Knowledge of an Adversary]  \label{def:knowledge_adversary}
	An adversary's knowledge encompasses probabilistic beliefs about all dataset entries, formalized as a prior distribution $p(x)$ over $\mathcal{X}$. 
\end{definition}

The distribution $p(x)$ is actually the adversary's a priori probabilities for which dataset is the true dataset, and represents his a priori knowledge of the situation \cite{DBLP:conf/icalp/Dwork06}. Similar notion is fundamental in cryptography \cite{DBLP:journals/jcss/GoldwasserM84}. 

We will abuse $X$ to name the adversary if $X \sim p(x)$. 
An adversary $X$ is said to be \emph{practical} if  an adversary can realistically possess knowledge $X$ in the real world,  excluding scenarios where the adversary is virtually omniscient in large-scale settings (e.g., $H(X) \approx 0$).

\begin{definition}[Probability Simplex \cite{DBLP:journals/fttcs/DworkR14}] \label{def:probability_simplex}
	For discrete set $B$, define $\Delta(B)$ as the simplex of probability distributions over $B$. Thus $\Delta(\mathcal{X})$ and $\Delta(\mathcal{X}_i)$ represent the probability spaces for complete datasets and individual records, respectively.
\end{definition}

We adopt notational shorthand where random variable $X$ may simultaneously denote both the variable and its distribution. For $X = (X_{I_1}, X_{I_2})$ following joint distribution $p(x_{I_1})p(x_{I_2}|x_{I_1})$, we write $p(x_{I_1}) \in \mathbb{P}$ and $p(x_{I_2}|x_{I_1}) \in \mathbb{P}$ if $p(x_{I_1})p(x_{I_2}|x_{I_1}) \in \mathbb{P}$ where $\mathbb P \subseteq \Delta(\mathcal{X})$.

\paragraph{Privacy Mechanism Formalism} For deterministic query function $f: \mathcal{X} \to \mathcal{Y}$ with finite range $\mathcal{Y}$, we define:

\begin{definition}[Privacy Mechanism]\label{def:privacy_mechanism}
	A \emph{privacy mechanism/channel} for $f$ is a stochastic channel $(\mathcal{X}, p(y|x), \mathcal{Y})$ represented by a probability transition matrix. The curator applies this mechanism to query results, generating randomized outputs $y \in \mathcal{Y}$ according to transition probabilities $p(y|x)$.
\end{definition}

\paragraph{Computational Model \cite{DBLP:journals/fttcs/DworkR14}}
In this paper, we assume the existence of a trusted curator who maintains the true dataset $x$. When a user submits a query function $f: \mathcal{X} \to \mathcal{Y}$ to the curator, the curator employs a privacy mechanism $p(y|x)$ to generate an output $y\in \mathcal Y$.  
This output $y$ is then returned to the user as a privacy-preserving approximation of the true result $f(x)$.

%This formulation separates deterministic query ($f$) from stochastic privacy protection ($p(y|x)$). The curator's role involves selecting appropriate mechanisms to sanitize query outputs before release.

\paragraph{Security Foundations} Current debates on privacy standards \cite{DBLP:journals/concurrency/0055ZL022} motivate our return to first principles. Dalenius' desideratum \cite{dalenius1977towards} – that database access should not increase individual record knowledge – finds natural expression in Shannon's perfect secrecy framework \cite{6769090}. While computational relaxations like semantic security \cite{DBLP:journals/jcss/GoldwasserM84} exist, their complexity makes information-theoretic approaches preferable for foundational analysis. Computational variants will be discussed in Section~\ref{sec-computational-variants}.

\begin{definition}[Dalenius Security]\label{def:dalenius_security}
	A privacy channel $(\mathcal{X}, p(y|x), \mathcal{Y})$ is called:
	\begin{itemize}
		\item $\epsilon$-\emph{perfectly Dalenius secure} if for any adversary $X \in \Delta(\mathcal{X})$ and all $i \in [n]$, we have $I(X_i;Y) \leq \epsilon$, where $Y$ is the channel output and $I(\cdot;\cdot)$ denotes mutual information.
		\item $\epsilon$-\emph{practically Dalenius secure} (or $\epsilon$-Dalenius secure) if the above holds for all \emph{practical} adversaries (e.g., $H(X) \gg 0$ for large-scale dataset).
	\end{itemize}
\end{definition}

The parameter $\epsilon$ is called the \emph{privacy budget} \cite{DBLP:journals/fttcs/DworkR14}. The ideal Dalenius objective corresponds to $\epsilon$-perfect security as $\epsilon \rightarrow 0$. However, we introduce the practical variant to enhance data utility, recognizing that perfect security often renders data unusable \cite{DBLP:conf/icalp/Dwork06}. This mirrors the relationship between perfect secrecy and semantic security in cryptography \cite{DBLP:conf/focs/Yao82a}, where computational hardness assumptions enable practical security guarantees.

\begin{remark}
	Our framework employs Shannon mutual information $I(X_i;Y)$ instead of max-mutual information $I_{\infty}(X_i;Y)$ (see Definition~\ref{def:max-MI}), despite the latter providing stronger guarantees since $I(X_i;Y) \leq I_{\infty}(X_i;Y)$. This choice is deliberate, as it strikes a balance between two considerations:
	\begin{enumerate}
		\item \textit{Improved utility for large $n$}: This is particularly relevant for real-world datasets, which often involve large sample sizes.
		\item \textit{Alignment with information theory foundations}: It enables seamless leveraging of established results and techniques from information theory to address data privacy challenges, as demonstrated in subsequent sections of this paper.
	\end{enumerate}
	Small dataset scenarios may warrant $I_{\infty}(X_i;Y)$ instead; see Section~\ref{subsec:related-work:other-measures} for detailed discussion.
\end{remark}

\begin{proposition}[Post-Processing Invariance]\label{prop:post-processing}
	Any $\epsilon$-Dalenius secure channel $(\mathcal{X}, p(y|x), \mathcal{Y})$ maintains its security under deterministic post-processing. Formally, for any function $g:\mathcal{Y} \to \mathcal{Z}$ and induced channel $(\mathcal{X}, p(z|x), \mathcal{Z})$, we have:
	\begin{align*}
		X \rightarrow Y \rightarrow g(Y) = Z \implies I(X_i;Z) \leq I(X_i;Y) \leq \epsilon.
	\end{align*}
\end{proposition}

\begin{proof}
	The Markov chain $X_i \rightarrow Y \rightarrow Z$ immediately satisfies the data processing inequality \cite[Theorem 2.8.1]{DBLP:books/daglib/0016881}. Thus $I(X_i;Z) \leq I(X_i;Y) \leq \epsilon$ by channel security.
\end{proof}

\begin{definition}[Individual Channel Capacity]  \label{def:individual_channel_capacity}
	For a function $f:\mathcal{X} \to \mathcal{Y}$ with privacy channel $p(y|x)$, define its \emph{individual channel capacity} relative to adversary class $\mathbb{P} \subseteq \Delta(\mathcal{X})$ as:
	\begin{align}
		C_1^{\mathbb{P}} = \max_{i \in [n],  X \in \mathbb{P}} I(X_i;Y).
	\end{align}
	The \emph{absolute capacity} $C_1$ corresponds to $\mathbb{P} = \Delta(\mathcal{X})$.
\end{definition}

The individual channel capacity quantifies the maximal information leakage of an individual to the adversary class $\mathbb P$ when the output $Y$ is observed. The adversary class $\mathbb P$ is used to model the class of practical adversaries to improve utility (e.g., $\mathbb P = \{p(x): H(X) \gg 0\}$ for large-scale dataset).
We will analyze it in detail in Section \ref{sec-individual-channel-capacity}, along with the discussion of $\mathbb P$ in Assumption \ref{assum:bounded_knowledge}.

\begin{definition}[Information Privacy]\label{def:information_privacy}
	A channel $p(y|x)$ satisfies $\epsilon$-\emph{information privacy} relative to $\mathbb{P}$ if $C_1^{\mathbb{P}} \leq \epsilon$. We omit $\mathbb{P}$ when $\mathbb{P} = \Delta(\mathcal{X})$, making $\epsilon$-information privacy equivalent to $\epsilon$-perfect Dalenius security.
\end{definition}

The channel guarantees that the maximal information leakage of an individual to an adversary class $\mathbb{P}$ is bounded by $\epsilon$ when observing output $Y$, provided that $Y$ is generated by an $\epsilon$-information private channel. This implies that the channel satisfies $\epsilon$-Dalenius security if $\mathbb P$ is a class of practical adversaries.

\begin{remark}
	The adversary class $\mathbb{P}$ operationalizes the privacy-utility tradeoff – tighter $\mathbb{P}$ constraints permit smaller $\epsilon$ while maintaining utility. We will discuss two kinds of $\mathbb P$ in Section \ref{subsec:dp-model} and Section \ref{subsec:new-model}, respectively.
\end{remark}

\subsection{The Differential Privacy Model} \label{subsec:dp-model}

A central debate in differential privacy concerns whether its security guarantees depend on the \emph{independence assumption} -- the premise that individuals' data records are mutually independent \cite{DBLP:journals/concurrency/0055ZL022}. This section establishes the necessity of this assumption for achieving Dalenius security in differential privacy. We first introduce a specialized information privacy framework and formalize the independence assumption, then demonstrate its equivalence to differential privacy under this constraint.

\begin{definition}[Max-Mutual Information \cite{DBLP:conf/focs/RogersRST16}] \label{def:max-MI}
	For random variables $ X $ and $ Y $, the max-mutual information is defined as
	\begin{align}
		I_{\infty}(X;Y) = \max_{x \in \mathcal{X}, y \in \mathcal{Y}} \log \frac{p(y|x)}{p(y)},
	\end{align}
	where mutual information $ I(X;Y) $ represents the expectation of $ \log \frac{p(y|x)}{p(y)} $.
\end{definition}

\begin{definition}[Max-Information Privacy] \label{def:max_information_privacy}
	A privacy channel $ p(y|x) $ satisfies $ \epsilon $-max-information privacy with respect to $ \mathbb{P}_{\text{ind}} $ if
	\begin{align}
		C_{1,\infty}^{\mathbb P_{\text{ind}}} :=  \max_{i \in [n], X \in \mathbb{P_{\text{ind}}}} I_{\infty}(X_i;Y) \leq \epsilon,
	\end{align}
	where $ \mathbb{P}_{\text{ind}} $ denotes the set of all product distributions over $ \mathcal{X} $:
	\begin{align}
		\mathbb{P}_{\text{ind}} = \left\{ X \in \Delta(\mathcal{X}) : p(x) = \prod_{i=1}^n p(x_i), \forall x = (x_1, \ldots, x_n) \in \mathcal{X} \right\}
	\end{align}
\end{definition}

\begin{assumption}[Independence Assumption] \label{assum:independence_assumption}
	Assume $ X \in \mathbb{P}_{\text{ind}} $, then the adversary $X$ is practical.
\end{assumption}

It should be noted that Assumption \ref{assum:independence_assumption} imposes overly stringent conditions that may not hold in practical scenarios.

\begin{definition}[Differential Privacy \cite{DBLP:journals/fttcs/DworkR14}] \label{definition-1}
	A privacy channel $ p(y|x) $ satisfies $ \epsilon $-differential privacy if for all $ y \in \mathcal{Y} $ and neighboring datasets $ x, x' \in \mathcal{X} $ ($ \|x - x'\|_1 \leq 1 $):
	\begin{align} \label{Eq:dp_inequality}
		\log \frac{p(y|x)}{p(y|x')} \leq \epsilon.
	\end{align}
\end{definition}

\begin{theorem} \label{proposition-1}
	Under Assumption \ref{assum:independence_assumption}, the following are equivalent:
	\begin{enumerate}
		\item $ p(y|x) $ satisfies $ \epsilon $-max-information privacy with respect to $ \mathbb{P}_{\text{ind}} $
		\item $ p(y|x) $ satisfies $ \epsilon $-differential privacy
	\end{enumerate}
\end{theorem}

\begin{proof}
	($ 2 \Rightarrow 1 $): For neighboring $ x, x' $ and any $ y $, differential privacy gives $ \log \frac{p(y|x)}{p(y|x')} \leq \epsilon $. For $ X \in \mathbb{P}_{\text{ind}} $:
	\begin{align}
		C_{1,\infty}^{\mathbb{P}_{\text{ind}}} = \max_{i, X \in \mathbb{P}_{\text{ind}}} I_{\infty}(X_i;Y) 
		&= \max_{i, x_i, y} \log \frac{\sum_{x_{-i}} p(y|x) p(x_{-i})}{\sum_{x_{-i}'} p(y|x') p(x_{-i}')} \\
		&\leq \epsilon. \quad \text{(by differential privacy)}
	\end{align}
	
	($ 1 \Rightarrow 2 $): For neighboring $ x = (x_i, x_{-i}) $, $ x' = (x_i', x_{-i}) $:
	\begin{align}
		\log \frac{p(y|x)}{p(y|x')} \leq \epsilon. \quad \text{(by max-information privacy)}
	\end{align}
\end{proof}

%m \ref{proposition-1} implies that any $ \epsilon $-differentially private channel is $ \epsilon $-Dalenius secure under Assumption \ref{assum:independence_assumption}. It shows that the (Dalenius) security of differential privacy relies on the independence assumption. 

%Theorem \ref{proposition-1} establishes that under Assumption \ref{assum:independence_assumption}, every $\epsilon$-differentially private channel satisfies $\epsilon$-Dalenius security. This result demonstrates the fundamental dependence of differential privacy's (Dalenius) security guarantees on the independence assumption.

Theorem \ref{proposition-1} implies that $\epsilon$-differential privacy is essentially equivalent to $\epsilon$-information privacy under the independence assumption.

\subsection{Enhanced Privacy Framework} \label{subsec:new-model}

Given the impracticality of Assumption \ref{assum:independence_assumption}, we propose a refined security assumption:

\begin{assumption}[Bounded Knowledge] \label{assum:bounded_knowledge}
	If $ X \in \Delta(\mathcal{X})_b $, then the adversary $X$ is practical, where
	\begin{align}
		\Delta(\mathcal{X})_b = \{ X \in \Delta(\mathcal{X}) : H(X) \geq b\}, 
	\end{align}
	with $b$ being a positive parameter and $H(X)$ being the entropy of $X$.
\end{assumption}

Entropy $H(X)$ quantifies the uncertainty in an adversary's knowledge $X$ regarding the secret dataset.
Assumption \ref{assum:bounded_knowledge} implies that, for sufficiently large datasets, no single adversary can possess complete information about all data within the secret dataset. This assumption is grounded in four key observations:
\begin{enumerate}
	\item \textbf{Partial Knowledge}: Real-world datasets typically contain non-overlapping information with public sources (e.g., Netflix vs IMDb data \cite{DBLP:conf/sp/NarayananS08}).
	\item \textbf{Scale Advantage}: Dataset entropy $ H(X) $ grows exponentially with size, unlike fixed-length cryptographic keys.
	\item \textbf{Long-Tail Distributions}: Prevalence of rare entries increases uncertainty.
	\item \textbf{Query Limitations}: Practical constraints prevent exhaustive dataset queries.
\end{enumerate}

We maintain that Assumption \ref{assum:bounded_knowledge} fundamentally distinguishes data privacy in \emph{large-scale datasets} from cryptographic systems with \emph{small secret keys}. While cryptographic hardness assumptions establish a satisfactory security-utility tradeoff \cite{DBLP:conf/focs/Yao82a}, our bounded knowledge assumption specifically governs the privacy-utility tradeoff in data privacy scenarios.

Under Assumption \ref{assum:bounded_knowledge}, $ \epsilon $-information private channels with respect to $ \Delta(\mathcal{X})_b $ are $ \epsilon $-Dalenius secure.

\begin{definition}[Balance Function] \label{definition-2}
	For a privacy channel with capacities $ C_1 = \max_{i \in [n], p(x) \in \Delta(\mathcal X)} I(X_i;Y) $ and $ C_1^b = \max_{i \in [n], p(x) \in \Delta(\mathcal X)_b} I(X_i;Y) $, the utility-privacy balance function is
	\begin{align}
		\delta = g(b) = C_1 - C_1^b, \quad b \in [0, \log |\mathcal{X}|].
	\end{align}
\end{definition}

\begin{lemma}[Monotonicity] \label{lemma-4}
	$ g(b) $ is non-decreasing with $ 0 = g(0) \leq g(b) \leq g(\log|\mathcal{X}|) \leq C_1 $.
\end{lemma}

\begin{proposition}[Privacy-Utility Tradeoff] \label{prop:privacy_utility_tradeoff}
	A channel satisfies $ \epsilon $-information privacy w.r.t. $ \Delta(\mathcal{X})_b $ iff it satisfies $ (\epsilon + g(b)) $-information privacy w.r.t. $ \Delta(\mathcal{X}) $.
\end{proposition}

Proposition \ref{prop:privacy_utility_tradeoff} establishes that under Assumption \ref{assum:bounded_knowledge}, the privacy-utility tradeoff can be optimized by constraining adversaries' knowledge to $H(X) \geq b$, which yields a reduction of $\delta$ in the privacy budget.

\begin{remark}[How to Choose $b$ in Practice?]
	The entropy constraint $H(X) \geq b$ ensures the adversary has at least $b$ bits of uncertainty about the dataset, where $b \in [0, \log|\mathcal{X}|]$. In practice, we aim to maximize $b$ to enhance data utility while maintaining the validity of Assumption \ref{assum:bounded_knowledge}. Setting $b$ too high risks violating this assumption if the adversary's actual knowledge exceeds the uncertainty threshold.
	
	To determine a suitable $b$ for a specific application, we estimate the adversary's prior knowledge about the dataset. Consider the Netflix dataset \cite{DBLP:conf/sp/NarayananS08} as an example: it has significant overlap with public sources like IMDb. Assuming 10\% of Netflix data is publicly available, and the dataset contains 100,480,507 ratings (each with at least two possible values), we have $\log|\mathcal{X}| \geq \log(2^{100,480,507}) \approx 100,480,507$ bits. Removing the 10\% known data leaves approximately 90,432,456 bits of uncertainty. Setting $b = \frac{3}{4} \times 90,432,456 \approx 67,824,342$ bits is therefore reasonable—it requires the adversary to obtain approximately 22,608,114 additional bits to violate Assumption \ref{assum:bounded_knowledge}, which is challenging when queries are protected by $\epsilon$-information privacy guarantees and query frequency is restricted as in differential privacy.
	
	This example demonstrates how domain knowledge about data overlaps and adversary capabilities can guide practical $b$ selection. The key is balancing utility optimization with realistic adversary modeling, ensuring the entropy constraint reflects actual knowledge gaps while providing meaningful privacy protection.
\end{remark}

\section{From Individual Channel Capacity to Channel Capacity} \label{sec-individual-channel-capacity}

The compatibility between information privacy and differential privacy models established in the preceding section raises a fundamental question: \emph{Can classical privacy channels---such as Laplace, Gaussian, and exponential mechanisms---be effectively employed within the information privacy framework?} To address this, we must first develop methodologies for evaluating individual channel capacities of privacy channels. This section presents a systematic approach to analyze these capacities through convex combinations of elementary channels.

We emphasize that calculating individual channel capacity differs fundamentally from solving the multi-access channel capacity problem in information theory. While the latter assumes independent senders \cite[\S 15]{DBLP:books/daglib/0016881}, our framework explicitly handles dependent individuals---a critical distinction that renders our problem significantly more complex.

\subsection{Illustrative Case Study} \label{subsec-simple-example}

To elucidate our methodology, we first analyze a concrete example before presenting general results. This pedagogical approach helps ground the abstract mathematical concepts in practical computation.

\begin{example} \label{example-1}
	Consider $\mathcal X_1 = \{0,1,2\}$ and $\mathcal X_2 = \{0,1\}$ with $\mathcal X = \mathcal X_1 \times \mathcal X_2$. Define the query function $f: \mathcal X \rightarrow \mathcal Y$ as:
	\begin{align}
		f(x_1,x_2) =  
		\begin{cases} 
			1 & x_1 = x_2 \\
			0 & \text{otherwise}
		\end{cases}
	\end{align}
	yielding $\mathcal Y = \{0,1\}$. The corresponding privacy channel's probability transition matrix is visualized in Fig. \ref{figure-1}.
	\begin{figure}[t]
		\centering
		\begin{tikzpicture}[every node/.style={anchor=north east,fill=gray!50,minimum width=1.4cm,minimum height=7mm},font=\footnotesize]
			\matrix (mA) [draw,matrix of math nodes]
			{
				p(y_2|x_{11},x_{22}) & p(y_2|x_{12},x_{22})  & p(y_2|x_{13},x_{22})  \\
				p(y_2|x_{11},x_{21}) & p(y_2|x_{12},x_{21})  & p(y_2|x_{13},x_{21})  \\
			};
			
			\matrix (mB) [draw,matrix of math nodes] at ($(mA.south west)+(3.5,-0.5)$)
			{
				p(y_1|x_{11},x_{22}) & p(y_1|x_{12},x_{22})  & p(y_1|x_{13},x_{22})  \\
				p(y_1|x_{11},x_{21}) & p(y_1|x_{12},x_{21})  & p(y_1|x_{13},x_{21})  \\
			};
			
			\draw[-](mA.north east)--(mB.north east);
			\draw[-](mA.north west)--(mB.north west);
			\draw[-](mA.south east)--(mB.south east);
			\draw[->]($(mB.south west)-(0.3,0.3)$)--($(mB.south east)-(0.3,0.3)$)node [near end, below right] {$x_1$};
			\draw[->]($(mB.south west)-(0.3,0.3)$)--($(mB.north west)-(0.3,0.3)$)node [near end, below left] {$x_2$};
			\draw[->]($(mB.north west)-(0.3,-0.3)$)--($(mA.north west)-(0.3,-0.3)$)node [near end, above left] {$y$};
		\end{tikzpicture}
		\caption{The probability transition matrix of the privacy channel $p(y|x)$ in Example~\ref{example-1}, with input domains $\mathcal{X}_1 = \{x_{11},x_{12},x_{13}\}$ for attribute $x_1$, $\mathcal{X}_2 = \{x_{21},x_{22}\}$ for attribute $x_2$, and output space $\mathcal{Y} = \{y_1,y_2\}$. The matrix elements specify conditional probabilities $p(y|x_1,x_2)$ for all $(x_1,x_2,y) \in \mathcal{X}_1 \times \mathcal{X}_2 \times \mathcal{Y}$.}\label{figure-1}
	\end{figure}
\end{example}

The individual channel capacity for this system can be expressed as:
\begin{align} \label{equation-54}
	C_1 = \max_{i\in [n], X\in \Delta(\mathcal X)} I(X_i;Y) &= \max_{i\in [n]} \max_{p(y|x_i) \in \mathbb Q_i} \max_{p(x_i) \in \Delta(\mathcal X_i)} I(p(y|x_i), p(x_i)) \\
	&= \max_{i\in [n]} \max_{p(y|x_i) \in \mathbb Q_i} C_{p(y|x_i)}
\end{align}
where 
\begin{align}
	\mathbb Q_i = \left\{ p(y|x_i) : p(x_{-i}|x_i) \in \Delta(\mathcal X_{-i}) \right\}
\end{align}
denotes the set of feasible transition matrices $p(y|x_i)$ induced by the privacy mechanism, $C_{p(y|x_i)}$ denotes the capacity of the transition matrix $p(y|x_i)$, and $\max_{p(y|x_i) \in \mathbb Q_i} C_{p(y|x_i)}$ represents the maximal amount of leaked information about the $i$th individual.

Recall that:
\begin{align}
	p(y|x_1) =& \sum_{x_{2}}p(x_{2}|x_1)p(y|x_1,x_{2}) \\
	=& (p(y|x_1,0),p(y|x_1,1))\begin{pmatrix}
		p(0|x_1) \\
		p(1|x_1)
	\end{pmatrix} \\
	=& p(0|x_1)(p(y|x_1,0),p(y|x_1,1))e_1 + p(1|x_1)(p(y|x_1,0),p(y|x_1,1))e_2 \\  
	=& p(0|x_1)p_1(y|x_1) + p(1|x_1)p_2(y|x_1),
\end{align}
where $e_1 = (1,0)^T$, $e_2 = (0,1)^T$ denote the two degenerate distributions (unit vectors) in $\Delta(\mathcal X_2)$, and
\begin{align}
	p_1(y|x_1) =& (p(y|x_1,0),p(y|x_1,1))e_1, \\
	p_2(y|x_1) =& (p(y|x_1,0),p(y|x_1,1))e_2.
\end{align}

Thus $p(y|x_1)$ is a convex combination of $p_1(y|x_1)$ and $p_2(y|x_1)$. Setting
\begin{align}
	\mathbb S_1 &= \left\{ p(y|x_1) : p(x_2|x_1) \in \Delta(\mathcal X_2)^{dg} \right\} = \{ p_1(y|x_1), p_2(y|x_1) \},
\end{align}
where $\Delta(\mathcal X_2)^{dg} = \{e_1,e_2\}$, we observe that any $p(y|x_1)$ in $\mathbb Q_1$ is a convex combination of elements in $\mathbb S_1$. Similarly, any $p(y|x_2)$ in $\mathbb Q_2$ is a convex combination of elements in $\mathbb S_2$, where
\begin{align}
	\mathbb S_2 = \left\{ p(y|x_2) : p(x_1|x_2) \in \Delta(\mathcal X_1)^{dg} \right\}.
\end{align}

Crucially, any $A \in \mathbb Q_i$ can be expressed as a convex combination of elements in $\mathbb S_i$, enabling capacity computation through basis channel analysis (Lemma \ref{lemma-6}).

\begin{lemma}[Convexity of Mutual Information \cite{DBLP:books/daglib/0016881}] \label{lemma-6}
	For a fixed input distribution $p(x_i)$, the mutual information $I(X_i;Y)$ is convex in the channel matrix $p(y|x_i)$.
\end{lemma}

\subsection{General Capacity Characterization} \label{subsec-general-result}

Extending our case study results, we present a general theorem for computing individual channel capacities in multi-dimensional privacy mechanisms.

\begin{theorem} \label{theorem-1}
	For a privacy channel $p(y|x)$ with $\mathcal X = \prod_{i=1}^n \mathcal X_i$, define:
	\begin{align}
		\mathbb Q_i &= \left\{ p(y|x_i) : p(x_{-i}|x_i) \in \Delta(\mathcal X_{-i}) \right\}, \\
		\mathbb S_i &= \left\{ p(y|x_i) : p(x_{-i}|x_i) \in \Delta(\mathcal X_{-i})^{dg} \right\}.
	\end{align}
	Then the individual channel capacity satisfies
	\begin{align}
		C_1 = \max_{i\in [n]} \max_{B \in \mathbb S_i} C_B,
	\end{align}
	where $C_B$ denotes the Shannon capacity of basis channel $B$.
\end{theorem}

\begin{proof}
	By construction, $\mathbb S_i$ contains extremal channels whose convex hull equals $\mathbb Q_i$. Lemma \ref{lemma-6} ensures that the maximum mutual information occurs at these extremal points. Computing over $\mathbb S_i$ rather than $\mathbb Q_i$ thus suffices while dramatically reducing search complexity from continuous to discrete spaces.
\end{proof}

This theorem establishes a fundamental bridge between privacy channel analysis and classical information theory. By decomposing complex privacy mechanisms into convex combinations of basis channels, we enable practical capacity calculations using well-established information-theoretic results. The next section demonstrates this methodology's power through applications to $\epsilon$-differential privacy mechanisms.

\section{Some Fundamental Privacy Channels} \label{sec:some-channels}

This section estimates individual channel capacities for key privacy channel archetypes using Theorem \ref{theorem-1}. We focus on three canonical privacy channels - the randomized response, Gaussian, and exponential channels - which serve as information-theoretic counterparts to differential privacy's core mechanisms. These channels exemplify distinct privacy protection paradigms while enabling closed-form capacity analysis.

\subsection{Binary Symmetric/Randomized Response Privacy Channel} \label{subsec-binary-channel}

The binary symmetric privacy channel generalizes both Shannon's binary symmetric channel \cite[\S 7.1.4]{DBLP:books/daglib/0016881} and Warner's randomized response mechanism \cite{DBLP:journals/fttcs/DworkR14}. For a query function $f: \mathcal{X} \rightarrow \mathcal{Y} = \{0,1\}$ as in Example \ref{example-1}, we define the privacy channel:
\begin{align} \label{equation-18}
	p(y|x) = (1-p)^{1-|y-f(x)|}p^{|y-f(x)|}, \quad x \in \mathcal{X}, y \in \mathcal{Y}
\end{align}
where $p \in (0,1)$ represents the bit-flip probability. This simple yet expressive model captures essential privacy-utility tradeoffs through symmetric error injection.

The individual channel capacity $C_1$ exhibits three regimes:

\textbf{Case 1 (One record):} When $n=1$ and $|\mathcal{X}|=2$, the channel reduces to a standard binary symmetric channel with capacity $C_1 = \log 2 - H(p)$, where $H(p) = -p\log p - (1-p)\log(1-p)$.

\textbf{Case 2 (Multiple records):} For $n \geq 2$, Theorem \ref{theorem-1} yields:
\begin{align}
	C_1 &= \max_{i\in[n]} \max_{p(y|x_i) \in \mathbb{S}_i} \max_{p(x_i)} \left( H(Y) - H(p) \right)
	\leq \log 2 - H(p). \label{equation-46}
\end{align}
If $|\mathcal{X}_i|=2$ for $ i \in [n]$, equality holds because $\mathbb{S}_i$ contains binary symmetric channels. When $|\mathcal{X}_i| \geq 3$ for some $i$, the upper bound \eqref{equation-46} persists though attainability requires further analysis.

The $\epsilon$-information privacy condition $\log 2 - H(p) \leq \epsilon$ translates to a symmetric noise interval $p \in (p^*, 1-p^*)$, where $p^*$ solves $H(p) = \log 2 - \epsilon$. This generates:

\begin{corollary}
	The binary symmetric channel \eqref{equation-18} satisfies $\epsilon$-information privacy over $\Delta(\mathcal{X})$ when $p \in (p^*,1-p^*)$, where $p^*$ solves $H(p^*) = \log 2 - \epsilon$.
\end{corollary}

\subsection{Column-Symmetric Privacy Channel} \label{subsec:column-symmetric}

Column-symmetric channels generalize differential privacy's additive noise mechanisms by enforcing permutation-invariant output distributions across inputs:

\begin{definition}[Column-Symmetric Channel]
	A privacy channel $p(y|x)$ for $f:\mathcal{X}\rightarrow\mathcal{Y}$ is column-symmetric if $\forall x,x'\in\mathcal{X}$, the two probability distributions $p(y|x)$ and $p(y|x')$ are permutations over $\mathcal{Y}$.
\end{definition}

This class includes (discrete) Laplace and Gaussian mechanisms. The capacity analysis proceeds as:
\begin{align}
	C_1 &\leq \log|\mathcal{Y}| - H(Z), \label{equation-14}
\end{align}
where $Z \sim p(y|x)$ for fixed $x$. Attaining equality requires weak symmetry:

\begin{definition}[Weakly Symmetric Channel] \label{definition-4}
	A column-symmetric channel is weakly symmetric if there exists an $i'\in[n]$ and $p^*(y|x_{i'})\in\mathbb{S}_{i'}$ with $\sum_{x_{i'}} p^*(y|x_{i'})$ constant across $y\in\mathcal{Y}$.
\end{definition}

\begin{corollary}
	For column-symmetric channels:
	\begin{align*}
		C_1 &\leq \log|\mathcal{Y}| - H(Z). 
	\end{align*}
	Equality holds, if weakly symmetric, achieved by uniform  $X_{i'}$.
	Thus $\epsilon$-information privacy holds when $H(Z) \geq \log|\mathcal{Y}| - \epsilon$.
\end{corollary}

\subsection{(Discrete) Exponential Privacy Channel} \label{subsec:exponential}

The exponential mechanism \cite{DBLP:journals/fttcs/DworkR14} is fundamental in the differential privacy model. 
We adapt the exponential mechanism to discrete outputs using distortion ranking:

\begin{definition}[Exponential Channel] \label{definition-7}
	For distortion ranking $\phi_x(y_j) = j-1$ where $d(y_{i_1},f(x)) \leq \cdots \leq d(y_{i_k},f(x))$, define:
	\begin{align} \label{equation-13}
		p(y|x) &= \frac{1}{\alpha} \exp\left(-\frac{\phi_x(y)}{N}\right), \quad \alpha = \sum_{i=0}^{k-1} \exp\left(-\frac{i}{N}\right).
	\end{align}
\end{definition}

The channel entropy derivation yields:
\begin{align}
	H(Z) = \log \frac{1-\exp(-k\lambda)}{1-\exp(-\lambda)}+ \frac{\lambda}{\exp(\lambda)-1}  -  \frac{k \lambda}{\exp(k\lambda)-1},  \label{equation-48}
\end{align}
where $\lambda = 1/N$. Solving $H(Z) \geq \log k - \epsilon$ numerically gives:

\begin{corollary}   \label{cor:exponential_mechanism}
	The exponential channel satisfies $\epsilon$-information privacy when $\lambda = 1/N$ solves $H(Z) \geq \log k - \epsilon$.
\end{corollary}

This analysis bridges exponential mechanisms with information-theoretic privacy, enabling parameter tuning via entropy constraints.

\subsection{Gaussian Privacy Channel}

\begin{definition}[Gaussian Privacy Channel] \label{def:gaussian}
	For continuous query $f:\mathcal{X} \rightarrow \mathbb{R}$ with bounded range $\mathcal{Y} \subseteq [-T,T]$, the Gaussian privacy channel adds noise:
	\begin{align}
		p(y|x) \sim \mathcal{N}(f(x), N),\quad N = \frac{T^2}{\exp(2\epsilon)-1}
	\end{align}
\end{definition}

\begin{corollary}[Channel Capacity] \label{cor:gaussian-cap}
	The individual channel capacity of the Gaussian privacy channel satisfies:
	\begin{align}
		C_1 = \frac{1}{2}\log\left(1 + \frac{T^2}{N}\right) \leq \epsilon
	\end{align}
	achieved when $N \geq \frac{T^2}{\exp(2\epsilon)-1}$. 
\end{corollary}

\section{Group and Composition Privacy}  \label{sec-group-composition-privacy}

We now turn to group privacy and composition privacy. \textbf{Group privacy} quantifies the privacy leakage of a specific group of individuals, capturing the collective information disclosure risk when multiple individuals' data is jointly considered. In contrast, \textbf{composition privacy} focuses on the cumulative privacy leakage of a single individual's data across multiple query outputs, addressing the risk of incremental information disclosure from multiple data access. Table \ref{tab:comparison_IP_group_composition} provides a concise comparison of information privacy, group privacy, composition privacy and analogous information-theoretic concepts, clarifying their core focuses and distinctions.

\subsection{Group Privacy} \label{sec-group-privacy}

\begin{table}[!b]
	%\caption{Comparison of Information Privacy, Group Privacy, Composition Privacy and Information-Theoretic Concepts}
	\caption{Comparison of Privacy and Analogous Information-Theoretic Concepts}
	\label{tab:comparison_IP_group_composition}
	\centering
	\footnotesize
	\begin{tabular}{@{} p{0.26\textwidth} p{0.43\textwidth} p{0.19\textwidth} @{}}
		\toprule
		\textbf{Concept} & \textbf{Core Meaning} & \textbf{Metric} \\
		\midrule
		Information Privacy & Information leakage of a single individual & $I(X_i;Y)$ \\
		\midrule
		Group Privacy       & Information leakage of a specific group of individuals (indexed by set $I$) & $I(X_I;Y)$ \\
		\midrule
		Composition Privacy & Cumulative leakage about an individual across multiple outputs $Y^1, \ldots, Y^k$ & $I(X_i;Y^1,\ldots,Y^k)$ \\
		\midrule
		Multi-access Channel & Multiple senders $\rightarrow$ Single receiver (Parallel to group privacy with dependent senders) &    Capacity region under independent senders\\
		\midrule
		Broadcast Channel & Single sender $\rightarrow$ Multiple receivers (Parallel to composition privacy)  & 
		Capacity region \\
		\bottomrule
	\end{tabular}
\end{table}

The group privacy problem investigates the collective privacy protection for groups of individuals within privacy-preserving mechanisms. Consider a scenario where a privacy channel's output $Y$ may reveal at most $\epsilon$ bits of information about any individual in Bob's family. A fundamental question arises: what is the maximum information disclosure about the entire family group through $Y$? 

\begin{definition}[Group Channel Capacity]
	For a function $f:\mathcal{X} \rightarrow \mathcal{Y}$ with privacy channel $p(y|x)$, we define its $k$-group channel capacity relative to $\mathbb{P} \subseteq \Delta(\mathcal{X})$ as:
	\begin{align}
		C_k^{\mathbb{P}} = \max_{\substack{X\in\mathbb{P} \\ I\subseteq[n]:|I|=k}} I(X_I;Y),
	\end{align}
	where $X_I = (X_{i_1},\ldots,X_{i_k})$ represents the random variables indexed by subset $I = \{i_1,\ldots,i_k\} \subseteq [n]$.
\end{definition}

\begin{definition}[Group Privacy]
	A privacy channel $p(y|x)$ for function $f:\mathcal{X} \rightarrow \mathcal{Y}$ satisfies $c$-group privacy relative to $\mathbb{P}$ if:
	\begin{align}
		\max_{k\in[n]} C_k^{\mathbb{P}} \leq k(\epsilon + c),
	\end{align}
	where $C_k^{\mathbb{P}}$ denotes the $k$-group channel capacity and $c$ represents a non-negative privacy parameter.
\end{definition}

\begin{lemma} \label{lemma-composition-2}
	Let $p(y|x)$ be a privacy channel for $f:\mathcal{X} \rightarrow \mathcal{Y}$. For any partition $I = I_1 \cup I_2$ with $I_1 \cap I_2 = \emptyset$, there exist $|\mathcal{X}_{I_1}|$ probability distributions $\{X^{x_{I_1}}:x_{I_1}\in\mathcal{X}_{I_1}\}$ over $\mathcal{X}$ satisfying:
	\begin{align}
		I(X_I;Y) = I(X_{I_1};Y) + \sum_{x_{I_1}} p(x_{I_1}) I(X_{I_2}^{x_{I_1}};Y^{x_{I_1}}),
	\end{align}
	where $X_{I_1} \sim p(x_{I_1})$, and $Y^{x_{I_1}}$ denotes the channel output when input follows $X^{x_{I_1}}$.
\end{lemma}

\begin{proof}
	We demonstrate the case for $I_1=\{1\}, I_2=\{2\}$, with generalization following similar reasoning. Applying the chain rule of mutual information:
	\begin{align}
		I(X_I;Y) &= I(X_1;Y) + I(X_2;Y|X_1) \\
		&= I(X_1;Y) + \sum_{x_1} p(x_1) I(X_2;Y|x_1).
	\end{align}
	
	For fixed $x_1 \in \mathcal X_1$, construct distribution $q(x) \in \Delta(\mathcal{X})$ with $q(x_1)=1$ and $q(x_{2}|x_1) = p(x_{2}|x_1)$. Let $\tilde{X} \sim q(x)$ and $\tilde{Y}$ be the corresponding output. Calculations show:
	\begin{align}
		q(y) = p(y|x_1), \quad q(x_2,y) = p(x_2,y|x_1), \quad q(x_2) = p(x_2|x_1)
	\end{align}
	leading to $I(\tilde{X}_2;\tilde{Y}) = I(X_2;Y|x_1)$. The result follows by substitution.
	\qed
\end{proof}

\begin{theorem} \label{theorem-composition-2}
	For a $(b,\delta)$-balanced privacy channel $p(y|x)$:
	\begin{itemize}
		\item Achieves $0$-group privacy over $\Delta(\mathcal{X})$
		\item Achieves $\delta$-group privacy over $\Delta(\mathcal{X})_b$
	\end{itemize}
\end{theorem}

\textbf{Comparison with Differential Privacy Group Privacy:} Our group privacy results exhibit both similarities and key differences with classical differential privacy. In differential privacy \cite{DBLP:journals/fttcs/DworkR14}, group privacy for $k$ individuals typically scales as $k\epsilon$, mirroring our linear scaling in Theorem \ref{theorem-composition-2}. However, differential privacy achieves this through the sequential application of the neighboring dataset definition, whereas our approach leverages information-theoretic chain rules and explicitly accounts for adversary knowledge constraints. Crucially, while differential privacy's group privacy guarantees depend on the independence assumption, our framework maintains group privacy under arbitrary data correlations through the bounded knowledge assumption ($H(X) \geq b$).

\subsection{Sequential Composition Privacy} \label{subsec:basic-composition-privacy}

Sequential composition privacy extends differential privacy's composition properties to information-theoretic frameworks. Consider a dataset subjected to multiple queries: how much cumulative information about an individual can adversaries extract from combined outputs?

\begin{definition}[Sequential Composition Privacy]
	For privacy channels $p(y^j|x)$ of $f_j:\mathcal{X} \rightarrow \mathcal{Y}^j$ ($j\in[m]$) respectively satisfying $\epsilon_j$-information privacy relative to $\mathbb{P}$, the composition channel $p(y|x) = \prod_{j=1}^m p(y^j|x)$ achieves $c$-composition privacy if:
	\begin{align}
		C_1^{\mathbb{P}} \leq \sum_{j=1}^m \epsilon_j + c,
	\end{align}
	where $C_1^{\mathbb{P}}$ is the individual channel capacity of the composed system.
\end{definition}

\begin{lemma} \label{lemma-composition-1}
	For $m=2$ queries, there exist $|\mathcal{Y}^2|$ distributions $\{X^{y^2}:y^2\in\mathcal{Y}^2\}$ such that:
	\begin{align}
		I(X_i;Y) = I(X_i;Y^2) + \sum_{y^2} p(y^2) I(X_i^{y^2};Y^{1,y^2}),
	\end{align}
	where $Y^{y^2}$ denotes composed outputs under input distribution $X^{y^2}$, and where $Y=(Y^1,Y^2)$.
\end{lemma}

\begin{proof}
	Following similar construction to Lemma \ref{lemma-composition-2}, define $q(x) = p(x|y^2)$ and compute:
	\begin{align}
		q(y^1) = p(y^1|y^2), \quad q(x_1,y^1) = p(x_1,y^1|y^2).
	\end{align}
	Mutual information calculations yield $I(\tilde{X}_1;\tilde{Y}^1) = I(X_1;Y^1|y^2)$, establishing the result.
	\qed
\end{proof}

\begin{table}[htb]
	\centering	\footnotesize
	\caption{Comparison of Composition Properties: Information Privacy vs. Differential Privacy}
	\label{tab:composition_comparison}
	\begin{tabular}{p{0.48\textwidth}p{0.48\textwidth}}
		\toprule
		\textbf{Information Privacy} & \textbf{Differential Privacy} \\
		\midrule
		Linear composition: $\sum \epsilon_j + c$ & Linear composition: $\sum \epsilon_j$ (basic) \\
		& Sublinear: $\mathcal{O}(\sqrt{m}\epsilon)$ (advanced) \\
		\addlinespace
		Privacy metric: Mutual information (average-case) & Privacy metric: Max-divergence (worst-case) \\
		\addlinespace
		Explicit adversary knowledge modeling ($H(X) \geq b$) & Adversary-agnostic \\
		\addlinespace
		Valid under arbitrary data correlations & Requires independence assumption for Dalenius security \\
		\addlinespace
		Utility improvement through $\delta$ terms & Fixed composition bounds \\
		\addlinespace
		Information-theoretic foundations & Divergence-based analysis \\
		\bottomrule
	\end{tabular}
\end{table}

\begin{theorem} \label{theorem-composition-1}
	For $(b,\delta^j)$-balanced channels $p(y^j|x)$ ($j\in[m]$):
	\begin{itemize}
		\item Achieve $0$-composition privacy over $\Delta(\mathcal{X})$
		\item Achieve $\sum_{j=1}^m \delta^j$-composition privacy over $\Delta(\mathcal{X})_b$
	\end{itemize}
\end{theorem}

\textbf{Comparison with Differential Privacy Composition:} Our composition results in Theorem \ref{theorem-composition-1} share the linear composition structure with differential privacy's basic composition theorem \cite{DBLP:journals/fttcs/DworkR14}, where the privacy parameters add up linearly ($\sum \epsilon_j$). However, there are fundamental differences in both the privacy metrics and the underlying assumptions. 

Differential privacy uses max-divergence and provides worst-case guarantees independent of the adversary's prior knowledge, leading to potentially conservative bounds. In contrast, our information-theoretic approach using mutual information provides average-case guarantees that can be tighter when adversary knowledge is bounded. Moreover, while differential privacy's advanced composition theorems \cite{DBLP:journals/fttcs/DworkR14} provide $\mathcal{O}(\sqrt{m}\epsilon)$ scaling for $m$ compositions under specific conditions, our framework offers the unique advantage of explicit utility improvements through the bounded knowledge assumption ($H(X) \geq b$), yielding additional $\delta^j$ terms that reduce the effective privacy cost.

Most importantly, our composition guarantees remain valid under arbitrary data correlations, whereas differential privacy's composition theorems rely on the independence assumption for their Dalenius security interpretation (as established in Theorem \ref{proposition-1}). This makes our composition framework particularly valuable for real-world datasets where correlations are inherent and the independence assumption is frequently violated.

\subsection{Comparative Analysis}

Table \ref{tab:composition_comparison} summarizes the key differences between our composition framework and classical differential privacy composition.

\section{Computational-Bounded Model} \label{sec-computational-variants}

While previous sections have analyzed information privacy under the assumption of unbounded computational power for both data owners and adversaries, real-world applications inherently operate under computational constraints. Recent advances in the computational complexity analysis of differential privacy \cite{DBLP:conf/tcc/BunCV16} motivate our development of a computational-bounded privacy framework. Building on Yao's computational information theory \cite{DBLP:conf/focs/Yao82a}, which leverages asymptotic analysis from computational complexity theory to handle large-scale inputs, we establish a rigorous foundation for modeling computational constraints in privacy-preserving systems.

\subsection{Formal Framework and Concrete Instantiation}

Following Yao's methodology \cite{DBLP:conf/focs/Yao82a}, we model random variables through source ensembles: each random variable $X_i$ is approximated by a source ensemble $\mathcal{S}_i$, and the joint random vector $X = (X_1,\ldots,X_n)$ by the product ensemble $\mathcal{S} = (\mathcal{S}_1,\ldots,\mathcal{S}_n)$. This framework enables computational approximations of classical information-theoretic measures, where the effective entropy $H_c(\mathcal{S})$ serves as the computational counterpart to Shannon entropy $H(X)$, and the effective mutual information $I_c(\mathcal{S}_i;\mathcal{T})$ approximates the standard mutual information $I(X_i;Y)$ for target ensemble $\mathcal{T}$.

\begin{example}[Practical Instantiation for Database Systems]
	Consider a medical database system where computational constraints naturally arise. The dataset $X$ represents patient records, and the privacy mechanism implements an information privacy query interface. In practice:
	\begin{itemize}
		\item \textbf{Adversary Constraints}: Real adversaries are computationally bounded—they cannot perform brute-force attacks on large datasets due to time and resource limitations.
		\item \textbf{Source Ensembles}: The medical records can be modeled as source ensembles where each patient's data $\mathcal{S}_i$ follows known statistical distributions based on population health data.
		\item \textbf{Effective Entropy}: $H_c(\mathcal{S})$ captures the computational difficulty of reconstructing complete patient information from partial knowledge and noisy queries.
	\end{itemize}
	This example demonstrates how our framework naturally accommodates real-world computational limitations that provide additional privacy protection.
\end{example}

The space of admissible source ensembles is characterized through entropy constraints. For any alphabet $\mathcal{X}$, we define $\Delta(\mathcal{X})_{b,c} = \{\mathcal{S}: H_c(\mathcal{S}) \geq b\}$ as the set of source ensembles with effective entropy at least $b$. This parameterized family of ensemble spaces forms the basis for our computational privacy definitions.

\subsection{Computational Privacy Foundations and Real-World Implications}

Central to our framework is the notion of computational information privacy. A privacy channel $p(y|x)$ implementing function $f:\mathcal{X} \rightarrow \mathcal{Y}$ satisfies $\epsilon$-computational information privacy with respect to adversary class $\mathbb{P}$ when its effective individual channel capacity $C_{1,c}^{\mathbb{P}} = \max_{\mathcal{S} \in \mathbb{P}} I_c(\mathcal{S}_i;\mathcal{T})$ is bounded by $\epsilon$, where $\mathbb{P} \subseteq \Delta(\mathcal{X})_{0,c}$ specifies admissible ensembles. This definition captures the worst-case information leakage under computational constraints.

\begin{example}[Medical Database System with Computational Constraints]
	\label{ex:medical-database}
	Consider a hospital database system that needs to release statistical queries while protecting patient privacy under realistic computational constraints. This scenario demonstrates how our computational-bounded framework naturally accommodates real-world limitations.
	
	\textbf{System Components:}
	\begin{itemize}
		\item \textbf{Dataset}: $n = 10^6$ patient records, where each record $X_i$ contains sensitive health information
		\item \textbf{Query}: Average cholesterol level computation via function $f: \mathcal{X} \rightarrow \mathbb{R}$
		\item \textbf{Data Curator}: Hospital database system with polynomial-time computational resources
		\item \textbf{Adversary Model}: Computationally bounded attacker (cannot break cryptographic primitives or perform brute-force attacks on large datasets)
	\end{itemize}
	
	\textbf{Computational Constraints Analysis:}
	\begin{itemize}
		\item \textbf{Theoretical Entropy}: Maximal dataset entropy $H(X) \approx 10^6 \times \log_2(|\mathcal{X}_i|)$ bits
		\item \textbf{Effective Entropy}: Due to computational limitations, adversary can only feasibly extract information from $\leq 10^2$ records
		\item \textbf{Constraint Setting}: Conservative bound $b = 0.9 \times 10^4 \times \log_2(|\mathcal{X}_i|)$ bits ensures Assumption \ref{assum:bounded_knowledge} holds
	\end{itemize}
	
	\textbf{Practical Implementation:}
	The hospital leverages this parameter $b$ to optimize their privacy mechanism:
	\begin{itemize}
		\item \textbf{Noise Reduction}: For Gaussian mechanism implementing average cholesterol query, noise scale can be reduced compared to information-theoretic setting
		\item \textbf{Enhanced Protection}: Computational constraints provide additional security layer
		\item \textbf{Utility Improvement}: System maintains $\epsilon$-computational information privacy relative to $\Delta(\mathcal{X})_{b,c}$ while achieving improved utility through balance function $\delta_c = g_c(b)$
	\end{itemize}
	
	\textbf{Utility-Privacy Tradeoff:}
	This approach demonstrates the practical advantage of our computational-bounded model: by explicitly modeling adversary computational limitations, we achieve stronger utility guarantees (reduced noise injection) while maintaining equivalent privacy protection for realistic threat scenarios.
\end{example}

The utility-privacy tradeoff is quantified through the computational balance function. For a given privacy channel $p(y|x)$, we analyze its effective channel capacities 
\begin{align}
	C_{1,c}^0 = \max_{\mathcal{S} \in \Delta(\mathcal{X})_{0,c}} I_c(\mathcal{S}_i;\mathcal{T}) \text{ and } C_{1,c}^b = \max_{\mathcal{S} \in \Delta(\mathcal{X})_{b,c}} I_c(\mathcal{S}_i;\mathcal{T}).
\end{align}
When there exists $\delta_c \geq 0$ satisfying $C_{1,c}^0 = C_{1,c}^b + \delta_c$, we say the channel is $(b,\delta_c)$-computationally balanced, with the balance function $\delta_c = g_c(b) = C_{1,c}^0 - C_{1,c}^b$ defined over $b \in [0,\log|\mathcal{X}|)$. This function suggests that the entropy constraint \( H_c(\mathcal{S}) \ge b \) can enhance the privacy budget to \( \epsilon + \delta_c \), thereby contributing to improved utility.

\subsection{Equivalence Results and System Design Implications}

The foundational equivalence between computational and information-theoretic privacy is established as follows. For a channel with balance function $\delta_c = g_c(b)$, $\epsilon$-computational information privacy relative to $\Delta(\mathcal{X})_{b,c}$ is equivalent to $(\epsilon+\delta_c)$-privacy under $\Delta(\mathcal{X})_{0,c}$, and further equivalent to $(\epsilon+\delta_c+\tau)$-information privacy in the standard model, where $\tau = C_1 - C_{1,c}^0$ captures the information-theoretic capacity gap.

Three significant implications emerge from this framework with direct relevance to system design:

\begin{enumerate}
	\item \textbf{Practical Security Guarantees}: Under standard cryptographic hardness assumptions and Assumption \ref{assum:bounded_knowledge}, $\epsilon$-computational information privacy directly implies $\epsilon$-Dalenius security. This provides a foundation for building practical systems that leverage computational hardness for enhanced utility.
	
	\item \textbf{Utility Optimization}: The utility advantage of the computational model is quantified by $\delta_c + \tau = C_1 - C_{1,c}^b \geq_a C_1 - C_1^b$, where the inequality follows from \cite[Theorem 2]{DBLP:conf/focs/Yao82a}. System designers can use this relationship to optimize the privacy-utility tradeoff by considering the actual computational capabilities of adversaries.
	
	\item \textbf{Unified Framework}: Special cases reveal deep connections to established cryptographic concepts: when $n=1$ with $\epsilon=0$ and $b=0$, our model recovers semantic security \cite{DBLP:journals/jcss/GoldwasserM84}; for $n=1$ and $b=0$, it coincides with computational differential privacy \cite{DBLP:conf/tcc/BunCV16}. This unification enables the design of systems that seamlessly integrate cryptographic and statistical privacy paradigms.
\end{enumerate}

\begin{example}[System Design Guidelines]
	For practitioners implementing privacy-preserving systems, our computational-bounded model suggests:
	\begin{itemize}
		\item \textbf{Resource-Based Parameter Tuning}: Adjust privacy parameters based on estimates of adversary computational resources. Systems facing resource-constrained adversaries can use larger $b$ while maintaining security.
		\item \textbf{Layered Defense}: Combine information-theoretic privacy mechanisms with computational hardness assumptions for defense in depth.
		\item \textbf{Adaptive Protection}: Dynamically adjust privacy parameters based on detected adversary capabilities and attack patterns.
	\end{itemize}
	These guidelines enable more efficient privacy protection in real-world systems where computational constraints provide natural security benefits.
\end{example}

\section{Evaluation}    \label{sec:evaluation}

This section presents an empirical evaluation of the proposed information privacy framework, focusing on two key aspects: 
(1) analyzing the relationship between individual channel capacity $C_1^b$ and entropy constraint $b$, and 
(2) comparing the utility-privacy tradeoffs of our framework against classical differential privacy mechanisms. 

While we recognize the importance of correlation-aware differential privacy variants such as Pufferfish \cite{DBLP:journals/tods/KiferM14} and Correlated differential privacy \cite{DBLP:journals/concurrency/0055ZL022}, we focus our comparison on classical differential privacy mechanisms due to their well-established mechanism designs and parameter tuning methodologies. A comprehensive comparison with these emerging frameworks would require substantial additional work to develop equivalent mechanism implementations and optimization strategies, which we defer to future research to ensure a fair and thorough evaluation.

Given the computational complexity of evaluating these quantities for general query functions, we use the parity function as a concrete and analytically tractable case study for our analysis.

\begin{example}
	\label{example-5} 
	Let $\mathcal X_i = \{0,1\}$ for $i \in [n]$. Consider the parity function defined as:
	\begin{align}
		f(x) = \begin{cases}
			0 & \text{ if } \sum_{i=1}^{n}x_i \pmod{2} = 0 \\
			1 & \text{ if } \sum_{i=1}^{n}x_i \pmod{2} = 1
		\end{cases}
	\end{align}
	for $x=(x_1,\ldots, x_n)\in\mathcal X$, with output space $\mathcal Y=\{0,1\}$. 
	
	The privacy mechanism is implemented as a binary symmetric channel: for each $x\in\mathcal X$ and $y\in \mathcal{Y}$,
	\begin{align}
		q(y|x) = \begin{cases}
			p    & \text{ if } y = 1 - f(x)\\
			1-p  & \text{ if } y = f(x) 
		\end{cases}
	\end{align}
	where $0 \le p \le 1/2$ represents the probability of a bit-flip error.
	
	The expected distortion is computed as:
	\begin{align}
		\mathbb E_{p_0} d(f(X),Y) = \sum_{x\in\mathcal X, y\in\mathcal Y} p_0(x)q(y|x)d(f(x),y),
	\end{align}
	where $p_0(x)$ is the uniform distribution over $\mathcal X$ and the distortion function is defined as:
	\begin{align}
		d(f(x),y) = \begin{cases}
			0 & \text{ if } f(x) = y \\
			1 & \text{ if } f(x) \ne y
		\end{cases}
	\end{align}
\end{example}

\begin{figure}[!htb]
	\centering
	\includegraphics[width=1\textwidth]{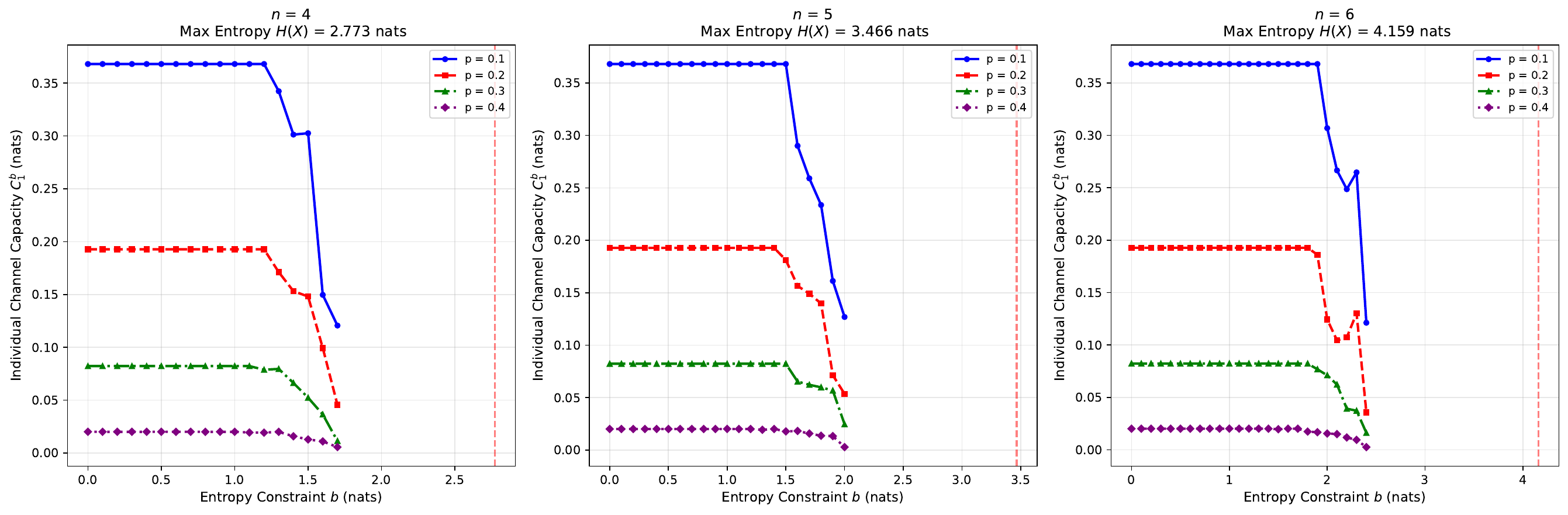}
	\caption{Variation of individual channel capacity $C_1^b$ with entropy constraint $b$, consisting of three subfigures corresponding to different numbers of records $n$ in the dataset: the left subfigure is for $n=4$, the middle subfigure for $n=5$, and the right subfigure for $n=6$. In each subfigure, there are four curves, which correspond to the binary symmetric-privacy channel with different bit-flip probabilities $p$ (representing the probability of flipping the query result to introduce randomness for privacy protection) as follows: the curve with the highest initial  capacity $C_1^b$ corresponds to $p=0.1$, the next to $p=0.2$, then $p=0.3$, and the curve with the lowest initial  capacity $C_1^b$ corresponds to $p=0.4$. 	
		The $x$-axis of each subfigure represents the entropy constraint $b$ (quantifying the lower bound of the adversary's knowledge uncertainty about the dataset, i.e., $H(X) \geq b$), and the $y$-axis denotes the individual channel capacity $C_1^b$ (measuring the maximum amount of information about any individual that an adversary can infer from the privacy channel output). All curves across the three subfigures consistently show that $C_1^b$ decreases as $b$ increases, confirming that the entropy constraint $H(X) \geq b$ effectively limits the potential information leakage of individual data by restricting the adversary's prior knowledge. Additionally, for the same $n$ and $b$, a larger $p$ leads to a smaller $C_1^b$, as higher randomness injection further reduces the adversary's ability to infer individual information. The non-monotonic segments observed in all curves originate from the non-convex nature of the $C_1^b$ calculation optimization problem: during the alternating optimization process (inspired by the Blahut-Arimoto algorithm), the algorithm converges to local optima temporarily, resulting in slight fluctuations in the information leakage value.}
	\label{fig:leakage-entropy-constraint}
\end{figure}

\subsection{Evaluating Individual Channel Capacity $C_1^b$}

The evaluation of individual channel capacity under entropy constraints requires solving the optimization problem:
\begin{subequations} \label{eq:optm-max_leakage}
	\begin{align} 
		C_1^b = & \quad \max_{i \in [n], p(x)}  I(X_i;Y) \\
		\text{s.t.} & \quad H(X) \geq b \label{eq:equation-4}
	\end{align}
\end{subequations}
where \( p(x) = p(x_i) p(x_{-i} | x_i) \), and the privacy mechanism \( q(y| x) \) is fixed. 

This constitutes a complex non-convex optimization problem. Our solution approach employs alternating algorithms inspired by the Blahut-Arimoto algorithm for channel capacity computation \cite{DBLP:journals/tit/Blahut72}. However, the introduction of entropy constraints and high-dimensional extensions presents significant additional challenges. Given the technical complexity and length of these algorithms, we focus here on the core privacy model construction and present experimental validation of key relationships.

Figure \ref{fig:leakage-entropy-constraint} illustrates how the individual channel capacity $C_1^b$ varies with the entropy constraint $b$ for different dataset sizes ($n=4,5,6$) and different bit-flip probabilities ($p=0.1,0.2,0.3,0.4$). The results demonstrate that $C_1^b$ decreases as $b$ increases across all configurations, confirming that the entropy constraint $H(X) \ge b$ effectively limits potential information leakage about any individual by constraining the adversary's prior knowledge. Additionally, for the same $n$ and $b$, a larger $p$ leads to a smaller $C_1^b$, as higher randomness injection further reduces the adversary's ability to infer individual information. The non-monotonic segments observed in all curves originate from the non-convex nature of the $C_1^b$ calculation optimization problem, where our algorithms converge to local optima temporarily during the alternating optimization process.

\subsection{Distortion Comparison to Differential Privacy Mechanisms}

We now compare the utility-privacy tradeoffs of our binary symmetric privacy channel against two fundamental differential privacy mechanisms: Laplace and Exponential mechanisms. For equitable comparison, we set the privacy parameter $\epsilon$ for each DP mechanism equal to the information leakage $C_1^b$ of our channel, ensuring equivalent privacy guarantees.

\begin{figure}[!hbt]
	\centering
	\includegraphics[width=1\textwidth]{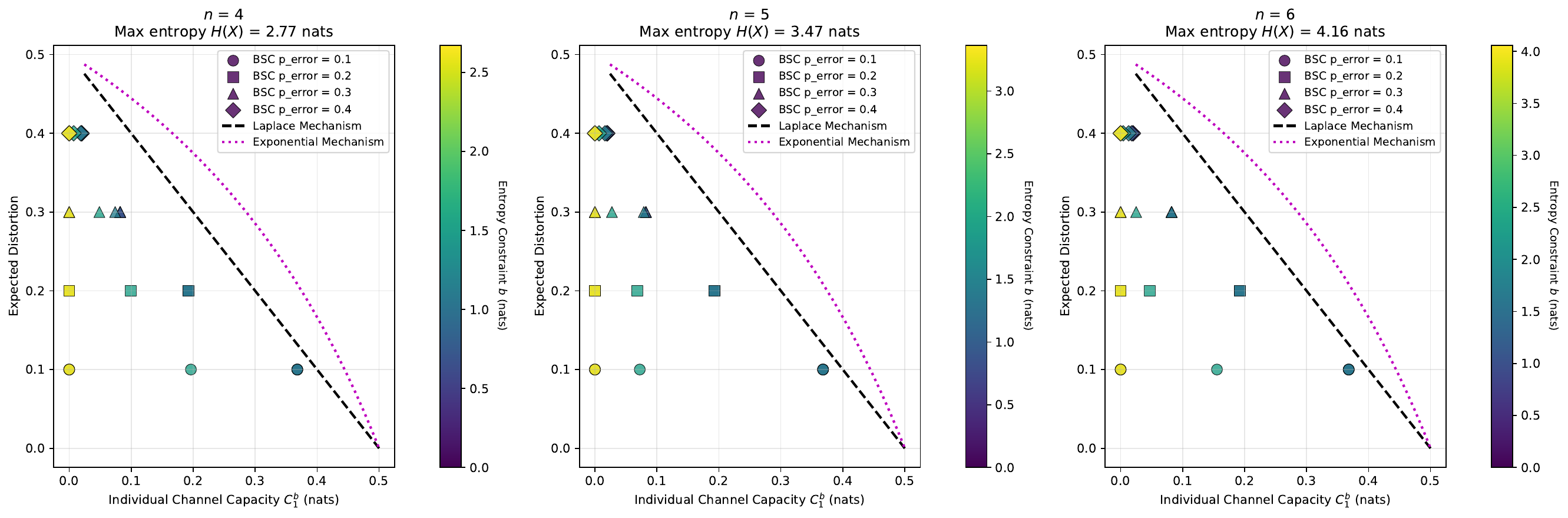}
	\caption{Comparison of individual channel capacity $C_1^b$ versus expected distortion $\mathbb E_{p_0} d(f(X),Y)$ across three privacy mechanisms for the binary parity query function, consisting of three subfigures corresponding to different numbers of records $n$ in the dataset: the left subfigure is for $n=4$ (with maximum entropy of 2.77 nats), the middle subfigure for $n=5$ (with maximum entropy of 3.47 nats), and the right subfigure for $n=6$ (with maximum entropy of 4.159 nats). In each subfigure, there are three kinds of curves, which correspond to different privacy mechanisms as follows: the four dotted curves with the lower privacy leakages at the same distortion level corresponds to the binary symmetric privacy channel, the next curve corresponds to the Laplace mechanism, and the curve with the highest privacy leakage corresponds to the Exponential mechanism. 
		The $x$-axis denotes the maximal privacy leakage/privacy budget, measured by the privacy parameter $\epsilon$: for the binary symmetric privacy channel, $\epsilon=C_1^b$ to ensure it satisfies $\epsilon$-Information Privacy with respect to the adversary class $\mathbb{P}_b$; for fair comparison, the privacy budget of the Laplace mechanism and Exponential mechanism  is set equal to $\epsilon$ to guarantee $\epsilon$-Differential Privacy.
		The $y$-axis of each subfigure represents the expected distortion $\mathbb E_{p_0} d(f(X),Y) $.
		All curves across the three subfigures consistently show two key results: (1) Under the same expected distortion, the binary symmetric privacy channel requires a smaller privacy budget $\epsilon$ (i.e., lower information leakage) than both the Laplace mechanism and the Exponential mechanism; (2) Increasing the entropy constraint $b$ further reduces the privacy leakage $C_1^b$ of the binary symmetric privacy channel at the same distortion level, which highlights the utility advantage of the information privacy framework enabled by the bounded knowledge assumption ($H(X) \geq b$).}
	
	\label{fig:leakage-distortion}
\end{figure}

\paragraph{Binary Symmetric Privacy Channel} 
For the parity function in Example \ref{example-5}, the binary symmetric privacy channel is defined as:
\begin{align} \label{equation-18}
	p(y|x) = (1-p)^{1-|y-f(x)|}p^{|y-f(x)|}, \quad x\in \mathcal X, y\in \mathcal Y.
\end{align}
The individual channel capacity with respect to $\Delta(\mathcal X)$ satisfies:
\begin{align} \label{eq:equation-5}
	C_1^0 = C_1 = \log 2 - H(p).
\end{align}
For specific parameter values:
\begin{itemize}
	\item $p = 0.1$: $C_1 \approx 0.531$ bits $\approx 0.368$ nats
	\item $p = 0.3$: $C_1 \approx 0.1187$ bits $\approx 0.08227657$ nats
\end{itemize}
The expected distortion simplifies to:
\begin{align}
	\mathbb E_{p_0} d(f(X),Y)  = \sum_{x\in\mathcal X} \frac{1}{|\mathcal X|} \times p \times 1 = p.
\end{align}

\paragraph{Laplace Mechanism}
The Laplace mechanism~\cite{DBLP:journals/fttcs/DworkR14} provides $\epsilon$-differential privacy for real-valued queries by adding calibrated Laplace noise. For the binary parity function $f: \mathcal{X} \rightarrow \{0,1\}$ with sensitivity $\Delta f = 1$:
\[
\mathcal{M}_L(x) = f(x) + \text{Laplace}\left(0, \frac{1}{\epsilon}\right).
\]
Since the output must be binary, we apply post-processing by thresholding:
\[
Y = \begin{cases} 
	1 & \text{if } \mathcal{M}_L(x) > 0.5 \\
	0 & \text{otherwise}
\end{cases}
\]
The expected distortion (error probability) is:
\[
\mathbb E_{p_0} d(f(X),Y)  = \frac{1}{2}\exp\left(-\frac{\epsilon}{2}\right).
\]

\begin{proof}
	When $f(x) = 0$, error occurs if Laplace noise $> 0.5$:
	\[
	P(\text{error} \mid f(x) = 0) = \int_{0.5}^{\infty} \frac{\epsilon}{2}\exp\left(-\epsilon|z|\right) dz = \frac{1}{2}\exp\left(-\frac{\epsilon}{2}\right).
	\]
	Similarly for $f(x) = 1$, error occurs if $1 + \text{Laplace}(0,1/\epsilon) \leq 0.5$. By symmetry, the overall error probability is $\frac{1}{2}\exp(-\epsilon/2)$.
\end{proof}

\paragraph{Exponential Mechanism}
The exponential mechanism~\cite{DBLP:journals/fttcs/DworkR14} provides $\epsilon$-differential privacy by sampling outputs proportional to their utility scores. For the binary parity function with utility function $u(x,y) = \mathbf{1}\{y = f(x)\}$ and sensitivity $\Delta u = 1$:
\[
P(Y = y \mid x) = \frac{\exp\left(\frac{\epsilon \cdot u(x,y)}{2}\right)}{\sum_{y' \in \{0,1\}} \exp\left(\frac{\epsilon \cdot u(x,y')}{2}\right)}.
\]
This simplifies to:
\[
P(Y = f(x) \mid x) = \frac{\exp(\epsilon/2)}{\exp(\epsilon/2) + 1}, \quad P(Y \neq f(x) \mid x) = \frac{1}{\exp(\epsilon/2) + 1}.
\]
The expected distortion is:
\[
\mathbb E_{p_0} d(f(X),Y)  = \frac{1}{\exp(\epsilon/2) + 1}.
\]

\paragraph{Distortion Comparison}
Figure \ref{fig:leakage-distortion} compares the individual channel capacity $C_1^b$ versus expected distortion across the three privacy mechanisms for different dataset sizes ($n=4,5,6$). The results demonstrate two key findings: (1) Under the same expected distortion, the binary symmetric privacy channel requires a smaller privacy budget $\epsilon$ (i.e., lower information leakage) than both the Laplace and Exponential mechanisms; (2) Increasing the entropy constraint $b$ further reduces the privacy leakage $C_1^b$ of the binary symmetric privacy channel at the same distortion level, which highlights the utility advantage of the information privacy framework enabled by the bounded knowledge assumption ($H(X) \geq b$).

\section{Related Works} \label{section:related-works}

\subsection{Differential Privacy, Perfect Secrecy and Semantic Security} \label{section:related-works:differential-privacy}

\begin{figure}[!t]
	\centering
	
	\begin{tikzpicture}[ 
		transform shape,  % Enable shape transformation
		scale=0.8,        % Reduce entire graphic by 30%
		node distance=10pt and 15pt,
		root/.style={rectangle, rounded corners, fill=red!10!white, draw=black, minimum width=3cm, minimum height=0.5cm,align=center, anchor=west, font=\small},
		main/.style={rectangle, rounded corners, fill=blue!10!white, draw=black, minimum width=3cm, minimum height=0.5cm,align=center, anchor=west, font=\footnotesize},
		leaf/.style={rectangle, rounded corners, fill=gray!10!white, draw=black, minimum width=1cm, minimum height=0.5cm,align=center, anchor=west, font=\scriptsize}
		]
		
		% Center node - Core concept
		\node (ipmodel0) [root] at (-8, 0) {IP Model($\epsilon, n$)};
		
		% Main branches
		\node (privacy1) [main] at (-5, 2) {Data Privacy($n>1$)};
		\node (cryptography1) [main] at (-5, -2) {Cryptography($n=1$)};
		
		% Curved connections to main branches
		\draw[thick, black, out=80, in=180] (ipmodel0) to (privacy1);
		\draw[thick, black, out=280, in=180] (ipmodel0) to (cryptography1);
		
		\node (perfect DS2_1) [leaf] at (-1,3) {Differential Privacy($\epsilon,n$) \\ Independence Assumption};
		\node (pd2) [leaf] at (-1,2){ \hspace{0.8cm} IP Model($\varepsilon, n$) \hspace{0.8cm} \\ Assumption \ref{assum:bounded_knowledge}};
		\node (pd3) [leaf] at (-1,1){Perfect Dalenius Security($\varepsilon, n$) \\ No Assumption };
		
		\draw[thick, black, out=0, in=180] (privacy1) to (perfect DS2_1);
		\draw[thick, black, out=0, in=180] (privacy1) to (pd2);
		\draw[thick, black, out=0, in=180] (privacy1) to (pd3);
		
		\node (pd2_1) [leaf] at (3.5, 2.5){ Perfect Dalenius Security \\ ($\epsilon,b=0$) };		
		\node (pd2_2) [leaf] at (3.5, 1.5){ Practical Dalenius Security \\ ($\epsilon,b>0$) };	
		
		\draw[thick, black, out=0, in=180] (pd2) to (pd2_1);			
		\draw[thick, black, out=0, in=180] (pd2) to (pd2_2);	
		
		\node (cryptography1_1) [leaf] at (-1, -1) {Perfect Secrecy ($\epsilon=0$) \\ No Assumption};				
		\node (cryptography1_2) [leaf] at (-1, -3) {Local Differential Privacy ($\epsilon>0$) \\ No Assumption};	
		\node (cryptography1_3) [leaf] at (-1, -2) {Semantic Security ($\epsilon=0$) \\ Difficult Assumptions};	
		
		\draw[thick, black, out=0, in=180] (cryptography1) to (cryptography1_1);
		\draw[thick, black, out=0, in=180] (cryptography1) to (cryptography1_2);	
		\draw[thick, black, out=0, in=180] (cryptography1) to (cryptography1_3);							
	\end{tikzpicture}
	
	% \caption{Conceptual Evolution and Relationships Between Privacy Frameworks and Encryption Models} 
	\caption{Conceptual Relationships Between Privacy and Security Notions} \label{figure-2}
\end{figure}

Figure \ref{figure-2} delineates the conceptual relationships among four critical privacy/security notions: the differential privacy, information privacy model, semantic security, and perfect secrecy. Our analysis reveals three fundamental connections.

First, the information privacy (IP) model fundamentally extends Shannon's classical encryption framework. The model comprises $n$ senders/individuals $X_1, \ldots, X_n$ and a receiver $Y$, with $\epsilon$-perfect Dalenius security defined as $\max_{i,X}I(X_i;Y) \le \epsilon$. Notably, when $n=1$ and $\epsilon=0$, this definition collapses precisely to Shannon's perfect secrecy \cite{6769090}. As demonstrated in Section \ref{sec-computational-variants}, under computational efficiency constraints for $X$ and $Y$, semantic security emerges as the computational counterpart to perfect secrecy, and by extension to $\epsilon$-perfect Dalenius security under standard cryptographic assumptions \cite{DBLP:conf/focs/Yao82a}.

Second, through the lens of independence assumption (Assumption \ref{assum:independence_assumption}), we establish that $\epsilon$-differential privacy becomes equivalent to $\epsilon$-Dalenius security when employing $I_{\infty}(X_i;Y)$ as the security metric. This equivalence, formalized in Theorem \ref{proposition-1}, finds corroboration in multiple foundational works \cite{DBLP:journals/tods/KiferM14}. For the special case of $n=1$, $\epsilon$-perfect Dalenius security aligns exactly with $\epsilon$-local differential privacy \cite{DBLP:journals/fttcs/DworkR14}.

Third, the security foundation of our proposed model in Section \ref{subsec:new-model} rests on Assumption \ref{assum:bounded_knowledge}. This assumption serves dual purposes: it replaces the independence assumption (Assumption \ref{assum:independence_assumption}) while mirroring the role of computational difficult assumptions in cryptography \cite{DBLP:conf/focs/Yao82a}. Although impractical in cryptographic contexts, this assumption demonstrates particular efficacy in data privacy protection scenarios.

The proposed framework inherits essential differential privacy properties including post-processing invariance, compositional guarantees, and group privacy preservation. Its practical superiority manifests particularly in complex data environments like graph analytics. Consider the graph data statistics domain \cite{li2023private}: traditional differential privacy implementations face a definitional dilemma between two neighborhood concepts:
\begin{description}
	\item[Edge neighboring datasets:] Differing by modification of a single edge
	\item[Node neighboring datasets:] Differing by complete alteration of a node's edge set
\end{description}
The edge neighborhood paradigm provides insufficient protection for high-degree nodes, while node neighborhood definitions impose excessive constraints that severely degrade utility. Our model circumvents this dichotomy by eliminating explicit neighborhood definitions, instead focusing on direct protection of data elements rather than data contributors. This approach inherently accommodates arbitrary inter-data correlations through its parameterization mechanism.

For graph analysis applications, our model requires only specification of either parameter $b$ (reflecting dataset scale and adversaries' knowledge) or privacy parameter $\delta$. In large-scale graph datasets (where $n \gg 1$), Assumption \ref{assum:bounded_knowledge} permits selection of larger $b$ values, thereby enabling enhanced utility through increased $\delta$ values without privacy compromise. 
Additionally, by accounting for high dependence in graph datasets, we can lower $b$ (or lower $\delta$) to ensure the practicality of Assumption \ref{assum:bounded_knowledge}. This flexibility is particularly advantageous for modern graph analysis tasks \cite{li2023private}.

\subsection{Information Theory} 

The information privacy model exhibits notable parallels with the multi-access channel framework in information theory \cite[\S 15]{DBLP:books/daglib/0016881}, particularly in their shared architecture of $n$ senders/individuals $X_1, \ldots, X_n$ transmitting to a common receiver $Y$. However, three fundamental distinctions exist between these frameworks:

\begin{enumerate}
	\item \textbf{Interference Handling}: In multi-access channels, senders must contend with both receiver noise and mutual interference, while operating under the critical independence assumption ($X_1, \ldots, X_n$ being independent). The information privacy model removes this independence constraint, explicitly addressing dependent senders as demonstrated in Theorems \ref{theorem-1} – a scenario outside conventional multi-access channel analysis.
	
	\item \textbf{Capacity Objectives}: Multi-access channel theory focuses on characterizing the channel capacity region, whereas our model prioritizes the evaluation of individual channel capacities under individuals' dependency constraint and adversaries' knowledge constraint $H(X) \ge b$.
	
	\item \textbf{Novel Constructs}: The balance function and individual channel capacity $C_1^b$ relative to $\Delta(\mathcal X)_b$ represent original contributions to information theory, introducing new analytical challenges in their computation and interpretation.
\end{enumerate}

These distinctions reveal fundamentally different problem spaces: multi-access channels emphasize capacity region analysis under independence assumptions, while our model focuses on dependent sender capacity evaluation with entropy constraint $H(X) \ge b$.

The group privacy and composition privacy problems further demonstrate connections to network information theory \cite[Ch.15]{DBLP:books/daglib/0016881}. Group privacy analysis parallels multi-access channel studies but maintains the aforementioned independence distinction. Composition privacy problems resemble hybrid network architectures combining multi-access and broadcast channels, differing primarily through our relaxation of independence assumptions.

While our proofs in Sections \ref{sec-individual-channel-capacity} --\ref{sec-group-composition-privacy} employ established information-theoretic methods, they provide significant advantages over existing privacy model analyses by systematically addressing dependency constraints that complicate proofs in prior work \cite{DBLP:journals/tods/KiferM14}.

Section \ref{sec-computational-variants} extends this theoretical foundation through adaptation of Yao's computational information theory \cite{DBLP:conf/focs/Yao82a}, demonstrating seamless integration with cryptographic security paradigms \cite{DBLP:journals/jcss/GoldwasserM84}.

Fundamentally, our framework revolves around capacity analysis for channel $(\mathcal X, p(y|x), \mathcal Y)$ and its sub-channels $(\mathcal X', p(y|x), \mathcal Y')$, where dependencies exist between both senders ($X_1,\ldots,X_n$) and receivers ($Y^1,\ldots,Y^m$) with $X=(X_1,\ldots,X_n), Y=(Y^1,\ldots,Y^m)$. This establishes deep theoretical connections between information theory and practical data privacy preservation.

Table \ref{tab:comparison_IP_group_composition} in Section \ref{sec-group-composition-privacy} provides a concise comparison of information privacy, group privacy, composition privacy and analogous  information-theoretic concepts, clarifying their core focuses and distinctions.

\subsection{Other Privacy Measures} \label{subsec:related-work:other-measures}

\begin{table}[!t]
	\caption{Comparison of Several Representative Privacy/Security Frameworks}
	\label{tab:comparison_privacy_models}
	\centering
	\scriptsize
	\begin{tabular}{@{} p{0.15\textwidth} p{0.2\textwidth} p{0.17\textwidth} p{0.15\textwidth} p{0.19\textwidth} @{}}
		\toprule
		\textbf{Framework}	 & \textbf{Metric} & \textbf{Assumption} & \textbf{Composition ($k$-queries)} & \textbf{Mechanism} \\
		\midrule
		Perfect Secrecy \cite{6769090}                            &  $I(X;Y) =0 $                        & No assumption to adversaries      &   $k \cdot I(X;Y) = 0 $            & One-time pad   \\
		\midrule
		Computational /Semantic Security \cite{DBLP:conf/focs/Yao82a}    &  $I_c(X;Y) \approx 0$  & Polynomial-time adversaries &  $k \cdot I_c(X;Y) \approx 0 $   &   RSA, AES \\
		\midrule
		Perfect Dalenius Security                   &  $I(X_i;Y) \le\epsilon $                        & No assumption to adversaries      &   $k\epsilon $                              & Exponential,  Gaussian  \\
		\midrule
		Differential Privacy \cite{DBLP:journals/fttcs/DworkR14}	& $I_{\infty}(X_i;Y) \le \epsilon$		& Independence Assumption & $k\epsilon$				& Exponential, Laplace \\
		\midrule
		Pufferfish \cite{DBLP:journals/tods/KiferM14}    &  $\log \frac{p(y \mid x, \theta)}{p(y \mid x', \theta)}  \leq \epsilon $  &   A class of adversaries $\mathbb P$   &     only proved special cases             &    only proved special cases              \\ 
		\midrule
		Correlated DP \cite{DBLP:journals/concurrency/0055ZL022}      & $I_{\infty}(X_i;Y) \le \epsilon$  & Independence Assumption + distribution of dataset     &  $k\epsilon$                     &  variant of Laplace                 \\
		\midrule
		Membership Privacy \cite{DBLP:conf/ccs/LiQSWY13}  &  $\log \frac{p(t \in \mathcal T|y)}{p(t)} \le \epsilon$  & A class of adversaries $\mathbb P$ &  Not proved    &  Not proved           \\ 
		\midrule
		CDP/RDP \cite{DBLP:conf/tcc/BunS16}   &  $D_{subG}(p(y|x) \Vert p(y|x'))$ $\preceq (\mu, \tau)$    & Independence Assumption     &     $(k\mu,\sqrt{k}\tau)$            &  Exponential,  Gaussian \\ 		
		\midrule
		Information Privacy        					&  $I(X_i;Y) \le \epsilon$             & Adversary class $H(X) \ge b$      &   $k\epsilon+k\delta$                      & Exponential,  Gaussian \\
		\midrule
		Computational Information Privacy        					&  $I_c(X_i;Y) \le \epsilon$             & Polynomial-time adversaries with $H_c(X) \ge b$      &   Not proved   & Not proved \\
		\bottomrule
	\end{tabular}
\end{table}

Beyond differential privacy, researchers have developed numerous alternative privacy frameworks \cite{zheng2025benchmarking}. Table \ref{tab:comparison_privacy_models} provides a comparative analysis of several representative privacy and security frameworks, examining them across key dimensions: privacy/security metrics, assumptions regarding adversaries' capabilities or knowledge, composition properties, and the mechanisms they enable.

\textbf{First}, preceding differential privacy, the $k$-anonymity model \cite{DBLP:journals/ijufks/Sweene02} emerged as the first extensively studied privacy paradigm. As noted in Section \ref{section-introduction}, $k$-anonymity and its derivatives - including $\ell$-diversity \cite{DBLP:conf/icde/MachanavajjhalaGKV06} and $t$-closeness \cite{DBLP:conf/icde/LiLV07} - fundamentally differ from differential privacy through their semantic dependence on data meaning, a characteristic absent in differential privacy. This semantic independence actually aligns with foundational security concepts like perfect secrecy and semantic security, a property inherited by both differential privacy and our proposed information privacy model as discussed in Section \ref{section:related-works:differential-privacy}.

\textbf{Second}, Concentrated Differential Privacy (CDP) \cite{DBLP:journals/corr/DworkR16} and R\'{e}nyi Differential Privacy (RDP) \cite{DBLP:conf/tcc/BunS16} employ privacy metrics based on Subgaussian Divergence and R\'{e}nyi divergence, respectively. These diverge from the KL divergence \( I(X_{i};Y) \) (used in Information Privacy (IP)) and the max divergence \( I_{\infty}(X_{i};Y) \) (underpinning standard Differential Privacy (DP)) \cite{DBLP:journals/corr/DworkR16}. CDP and RDP provide enhanced utility over standard DP, primarily due to their tighter advanced composition properties. These properties substantially reduce the cumulative privacy budget expenditure during iterative computations, making CDP and RDP particularly advantageous for privacy-preserving machine learning algorithms \cite{whitehouse2023fully} by enabling more complex computations under a fixed privacy guarantee.

However, like DP, both CDP and RDP focus exclusively on bounding individual privacy loss via neighboring dataset analyses, neglecting inference risks stemming from inherent data correlations. Consequently, they inherit the security limitations of DP and largely rely on the often unrealistic independence assumption.

In contrast, this paper focuses on defining a suitable adversary class (e.g., adversaries with bounded prior knowledge, \( H(X)\geq b \)) to achieve a superior privacy-utility tradeoff. We adopt KL divergence \( I(X_{i};Y) \), rather than max divergence \( I_{\infty}(X_{i};Y) \), as the privacy metric for two main reasons (see Section \ref{section:prelim}): (i) it yields better utility under our framework, and (ii) it aligns with information-theoretic principles, facilitating the adaptation of methods and results from information theory. While our approach is based on KL divergence, exploring other divergences tailored to the adversary class \( H(X)\geq b \) remains an open research question.

Another relaxation, bounded-leakage differential privacy \cite{DBLP:conf/forc/LigettPR20}, maintains the neighboring dataset paradigm and therefore largely inherits security dependencies on independence assumptions.

\textbf{Third}, the Pufferfish framework \cite{DBLP:journals/tods/KiferM14} defines its privacy metric as a bound on the \emph{difference} of two log-likelihood ratios:
\begin{align}
	\log \frac{p(y \mid x, \theta)}{p(y \mid x', \theta)}  \leq \epsilon,
	\label{Eq:pufferfish_definition}
\end{align}
where $x, x' $ can represent potential datasets, $\theta \in \mathbb{P}$ models the adversary's prior knowledge, and $\mathbb{P}$ denotes a class of admissible adversaries. While structurally reminiscent of differential privacy's bound (\ref{Eq:dp_inequality}), this formulation explicitly incorporates dataset correlations through the adversary's knowledge $\theta$.
Key distinctions and relationships of the Pufferfish and the information privacy frameworks are:
\begin{enumerate}
	\item \textbf{Adversary Knowledge Modeling:}
	\begin{itemize}
		\item The information-theoretic constraint $H(X) \geq b$ (Assumption \ref{assum:bounded_knowledge}) constitutes a specific instantiation of the adversary class $\mathbb{P}$ in the Pufferfish framework.
		\item The parameter $b$ in $H(X) \geq b$ provides sufficient flexibility to model varying adversary strengths: a \emph{larger $b$} encodes \emph{weaker} adversaries (such as implying lower intrinsic dataset correlation), while a \emph{smaller $b$} encodes \emph{stronger} adversaries (such as implying higher correlation, e.g., in graph-structured data). This parametric tuning effectively captures diverse adversarial scenarios.
		\item Consequently, while $\mathbb{P}$ offers greater \emph{theoretical generality} for adversary specification, the entropy-based constraint $H(X) \geq b$ provides a highly \emph{practical} and \emph{expressive} alternative without sacrificing representational adequacy for real-world applications.
	\end{itemize}
	
	\item \textbf{Privacy Metric:} Although both frameworks quantify privacy leakage, the mutual information of our approach (with $I(X_i;Y) \le\epsilon$) offers a distinct advantage: it enables the direct application of tools and theoretical results from \emph{information theory} to address privacy challenges in data analysis. This integration facilitates novel analyses and solution strategies as shown in this paper.
	
	\item \textbf{Composition and Mechanisms \cite{DBLP:journals/tods/KiferM14}:} 
	\begin{itemize}
		\item The Pufferfish framework's flexibility impedes derivation of general composition properties, with existing results primarily limited to specialized cases \cite{DBLP:journals/tods/KiferM14}.
		\item Key mechanisms including Gaussian mechanism (Corollary \ref{cor:gaussian-cap}) and Exponential mechanism (Corollary \ref{cor:exponential_mechanism}) lack general proofs within the framework.
		\item These limitations constrain Pufferfish's applicability in complex, multi-stage privacy scenarios.
	\end{itemize}
\end{enumerate}
     
Similar to Pufferfish and our framework, Membership privacy \cite{DBLP:conf/ccs/LiQSWY13} integrates adversary knowledge constraints to explicitly account for dataset correlations while optimizing utility. Its privacy metric bounds the max-information
\begin{align} 
	\log \frac{p(t \in \mathcal{T} \mid y)}{p(t)} \leq \epsilon,
	\label{Eq:membership_privacy}
\end{align}
which quantifies the risk of inferring whether an individual's data point $t$ belongs to the dataset.

Our entropy-based constraint $H(X) \geq b$ (Assumption \ref{assum:bounded_knowledge}) similarly specializes the adversary knowledge class, analogous to its relationship with Pufferfish. However, key distinctions emerge:
\begin{itemize}
	\item \textbf{Privacy Objective:} Membership privacy addresses \emph{membership inference} ($t \in \mathcal{T}$), while our framework protects against \emph{general data reconstruction}.
	\item \textbf{Metric Foundation:} The max-information metric in (\ref{Eq:membership_privacy}) represents a specialized case, whereas our mutual information bound $I(X_i;Y) \leq \epsilon$ provides a comprehensive information-theoretic foundation.
	\item \textbf{Theoretical Limitations:} Like Pufferfish, membership privacy lacks established composition theorems and general mechanism analyses (Gaussian, Laplace, Exponential) comparable to our framework's proven results (Theorem \ref{theorem-composition-1}, Corollaries \ref{cor:gaussian-cap}, \ref{cor:exponential_mechanism}), limiting its applicability in multi-stage data releases.
\end{itemize}

\textbf{Fourth}, Correlated Differential Privacy~\cite{DBLP:journals/concurrency/0055ZL022} or Dependent Differential Privacy~\cite{DBLP:conf/ndss/LiuMC16} extends differential privacy to incorporate dataset correlations, thereby enhancing resilience against correlation-based inference attacks. The two variants of DP model correlations as \emph{fixed, inherent properties} of the dataset \cite{DBLP:journals/concurrency/0055ZL022}. This core assumption fundamentally diverges from our framework's approach to correlations:
\begin{itemize}
	\item \textbf{Correlation Perspective:} Correlated DP and Dependent DP view correlations as static dataset attributes, whereas we adopt Shannon's cryptographic perspective \cite{6769090} where correlations constitute \emph{dynamic adversarial knowledge} about the secret dataset.
	\item \textbf{Adversary Modeling:} Under our framework, adversaries possess evolving distributional knowledge (including correlations) that may change over time—consistent with real-world threat models in cryptography.
\end{itemize}
This philosophical distinction yields critical analytical implications: while Correlated DP and Dependent DP strengthen DP against specific fixed correlations, our approach provides a unified framework for privacy analysis under arbitrary, time-varying adversarial knowledge.

\textbf{Fifth}, numerous privacy measures employ information-theoretic quantities like $I_{\infty}(X_i;Y)$ or $I(X_i;Y)$ \cite{DBLP:conf/ccs/LiQSWY13}, with many applying these metrics to analyze differential privacy \cite{DBLP:conf/ccs/CuffY16}. These approaches typically constrain adversarial knowledge to improve utility \cite{DBLP:journals/popets/DesfontainesP20}, motivating our Assumptions \ref{assum:bounded_knowledge} and \ref{assum:independence_assumption}.

Notably, our work diverges from prior mutual information applications in differential privacy analysis. While \cite{DBLP:conf/ccs/CuffY16} examines conditional mutual information $I(X_i;Y|X_{-i})$ and \cite{DBLP:journals/tit/WangYZ16} analyzes $I(X;Y)$, our Theorem \ref{proposition-1} establishes a critical equivalence condition revealing differential privacy's dependence on independence assumptions for Dalenius security.

\section{Conclusion} \label{section:conclusion}

We present a theoretically grounded privacy framework achieving Dalenius' vision by reconciling information-theoretic principles with practical data processing constraints. Its core innovation distinguishes cryptographic security (relying on small keys) from privacy-preserving data processing (handling massive datasets), enabling the replacement of differential privacy's problematic independence assumption with practical partial knowledge constraints. This framework delivers four significant advances: (1) Dalenius-compliant privacy resilient to correlated data attacks; (2) enhanced utility-privacy tradeoffs through adjustable partial knowledge constraints compared to perfect Dalenius security; (3) formal connections between foundational privacy mechanisms and information theory; and (4) empirical validation demonstrating superior utility-privacy tradeoffs against classical differential privacy mechanisms under equivalent privacy guarantees.

Key limitations of our current work center on the precise characterization of the privacy-utility balance function, $\delta = g(b)$, which is critical for optimizing parametric tradeoffs between privacy and data usefulness. To address this challenge and advance the framework, future research will pursue the following specific directions:

\begin{enumerate}
	\item \textbf{Numerical Characterization of \( g(b) \):} Building upon our empirical findings, a primary objective is to develop a comprehensive numerical characterization of the balance function. This will enable rigorous quantitative comparisons of privacy (quantified as information leakage) and utility (measured by distortion or accuracy) against established frameworks like differential privacy and Pufferfish. We plan to address this complex, high-dimensional optimization problem in a separate, forthcoming paper.
	\begin{itemize}
		\item Conduct \textbf{quantitative comparisons} with correlated differential privacy and Pufferfish privacy, evaluating their performance under varying correlation strengths and adversarial knowledge models.
		\item Systematically \textbf{evaluate utility-privacy tradeoffs} using standard metrics—such as distortion and query accuracy—to empirically demonstrate the advantages of our proposed framework.
	\end{itemize}
	
	\item \textbf{Empirical Validation on Diverse Benchmarks:} We will undertake extensive empirical validation to solidify the framework's practical relevance.
	\begin{itemize}
		\item Implement \textbf{synthetic and real-world experiments} to assess the framework's performance on common query types, including counting, averaging, and machine learning tasks.
		\item Incorporate \textbf{targeted case studies} on graph datasets and other high-correlation scenarios to rigorously test its resilience against sophisticated correlation-based inference attacks.
		\item Develop \textbf{practical methodologies for selecting the parameter \( b \)} based on observable dataset characteristics, such as scale, inherent correlation structure, and realistic estimates of adversarial prior knowledge.
	\end{itemize}
	
	\item \textbf{Extensions to Advanced and Sequential Applications:} To broaden the framework's applicability, we will explore its integration with modern data processing paradigms.
	\begin{itemize}
		\item Derive \textbf{tighter composition theorems} tailored for iterative algorithms—such as those used in federated learning and stochastic gradient descent—to enhance utility in multi-query and iterative settings.
		\item Adapt the framework to \textbf{sequential data processing paradigms}, including convolutional neural networks and transformer architectures, by incorporating temporal or structural constraints into the privacy analysis.
		\item Investigate the \textbf{fundamental interconnections} between computational information privacy and cryptographic techniques, aiming to unify security and privacy guarantees within a single, cohesive model.
	\end{itemize}
\end{enumerate}

Resolving these challenges promises significant advances for both the theoretical foundations and practical deployment of the information privacy framework. We posit that established techniques and emerging results from information theory will prove instrumental in addressing these open questions.

\section*{Declaration of Generative AI and AI-Assisted Technologies}
This manuscript represents an enhanced version of our previous preprint \cite{wu2021achievingdalenius}, where DeepSeek's AI technology has been employed to improve linguistic clarity and readability. Following the AI-assisted refinement, the authors have conducted thorough reviews and made necessary editorial revisions to the content. The authors assume full responsibility for all aspects of the published work, including but not limited to its academic integrity and scholarly content.

\section*{Acknowledgements}

We gratefully acknowledge the support and encouragement of numerous colleagues. In particular, the first author extends sincere thanks to collaborators and associates at the Institute of Software Chinese Academy of Sciences (ISCAS), Chongqing University of Technology (CQUT), and Lanzhou University of Finance and Economics (LZUFE). Special appreciation is also extended to family members for their invaluable and steadfast support throughout the duration of this research.

The authors are deeply grateful to the anonymous reviewers and the editorial team for their insightful comments, constructive feedback, and patient guidance. Their expertise has profoundly enhanced the overall quality, clarity, and scholarly rigour of this manuscript.

\section*{Funding}
This work was supported by the following grants: 
the Gansu Provincial Science and Technology Major Project - Industrial Category [22ZD6GA047],
the Gansu Provincial Higher Education Youth Doctoral Support Program [2023QB-070], 
and the University-level Research Project of Lanzhou University of Finance and Economics [Lzufe2021B-014], 
for the projects ``Research on the Application of Artificial Intelligence in Tumor Imaging Diagnosis and Treatment",
``Research on Key Issues in Information-Theoretic Foundations of Data Privacy", 
and ``Research on Key Issues in Privacy-Preserving Federated Learning", respectively.

\bibliographystyle{elsarticle-num} 
%\input{bib-files.tex}
%\bibliographystyle{unsrt}


%%       \cite{<label>} ==> [1]

%%
%Example citation, See \cite{lamport94}.


\begin{thebibliography}{99}
	\expandafter\ifx\csname url\endcsname\relax
	\def\url#1{\texttt{#1}}\fi
	\expandafter\ifx\csname urlprefix\endcsname\relax\def\urlprefix{URL }\fi
	\expandafter\ifx\csname href\endcsname\relax
	\def\href#1#2{#2} \def\path#1{#1}\fi
	
	\bibitem{2023arXiv230602781G}
	R.~{Gozalo-Brizuela}, E.~C. {Garrido-Merch{\'a}n}, {A survey of Generative AI
		Applications}, arXiv e-prints (2023) arXiv:2306.02781%\href
	%{http://arxiv.org/abs/2306.02781} {\path{arXiv:2306.02781}}, %\href
	%{https://doi.org/10.48550/arXiv.2306.02781}
	%{\path{doi:10.48550/arXiv.2306.02781}}.
	
	\bibitem{mayer2013big}
	V.~Mayer-Sch{\"o}nberger, K.~Cukier, Big Data: A Revolution That Will Transform
	How We Live, Work, and Think, Houghton Mifflin Harcourt, Boston, MA, 2013.
	
	\bibitem{DBLP:conf/sp/NarayananS08}
	A.~Narayanan, V.~Shmatikov, %\href{http://dx.doi.org/10.1109/SP.2008.33}
	{Robust de-anonymization of large sparse datasets}, 
	in: 2008 {IEEE} Symposium on Security and Privacy (S{\&}P 2008), 18-21 May 2008, Oakland, California, {USA}, 2008, pp. 111--125.
	% \newblock %\href {https://doi.org/10.1109/SP.2008.33}
	% {\path{doi:10.1109/SP.2008.33}}.
	% \newline\urlprefix\url{http://dx.doi.org/10.1109/SP.2008.33}
	
	\bibitem{carlini2021extracting}
	N.~Carlini, et~al., Extracting training data from large language models, in:
	USENIX Security Symposium, 2021, pp. 2633--2650.
	
	\bibitem{lee2022deduplicating}
	K.~Lee, D.~Ippolito, A.~Nystrom, C.~Zhang, D.~Eck, C.~Callison-Burch,
	N.~Carlini, Deduplicating training data makes language models better, in:
	Annual Meeting of the Association for Computational Linguistics (ACL 2022),
	Dublin, Ireland, May 22-27, 2022, 2022, pp. 8424--8445.
	% \newblock %\href {https://doi.org/10.18653/v1/2022.acl-long.577}
	% {\path{doi:10.18653/v1/2022.acl-long.577}}.
	
	\bibitem{hu2022membership}
	H.~Hu, Z.~Salcic, L.~Sun, G.~Dobbie, P.~S. Yu, X.~Zhang, Membership inference
	attacks on machine learning: A survey, ACM Computing Surveys (ACM Comput.
	Surv.) 54~(11s) (2022) 235:1--235:37.
	% \newblock %\href {https://doi.org/10.1145/3523273} {\path{doi:10.1145/3523273}}.
	
	\bibitem{voigt2017eu}
	P.~Voigt, A.~Von~dem Bussche, The EU GDPR: A Practical Guide, Springer, Berlin,
	Germany, 2017.
	
	\bibitem{europeanunion2024artificial}
	{European Union}, Artificial intelligence act, Official Journal of the EU
	(2024).
	
	\bibitem{near2025guidelines}
	J.~P. Near, D.~Darais, N.~Lefkovitz, G.~S. Howarth, Guidelines for evaluating
	differential privacy guarantees, Special Publication (SP) 800-226, National
	Institute of Standards and Technology (NIST) (March 2025).
	% \newblock %\href {https://doi.org/10.6028/NIST.SP.800-226}
	% {\path{doi:10.6028/NIST.SP.800-226}}.
	
	\bibitem{DBLP:journals/fttcs/DworkR14}
	C.~Dwork, A.~Roth, %\href{http://dx.doi.org/10.1561/0400000042}
	{The algorithmic foundations of differential privacy}, 
	Foundations and Trends in Theoretical Computer Science 9~(3-4) (2014) 211--407.
	% \newblock %\href {https://doi.org/10.1561/0400000042}
	% {\path{doi:10.1561/0400000042}}.
	% \newline\urlprefix\url{http://dx.doi.org/10.1561/0400000042}
	
	\bibitem{Shannon1948}
	C.~E. Shannon, %\href{https://ieeexplore.ieee.org/document/6773024/}
	{A mathematical theory of communication}, 
	The Bell System Technical Journal 27~(3) (1948) 379--423.
	% \newblock %\href {https://doi.org/10.1002/j.1538-7305.1948.tb01338.x}
	% {\path{doi:10.1002/j.1538-7305.1948.tb01338.x}}.
	% \newline\urlprefix\url{https://ieeexplore.ieee.org/document/6773024/}
	
	\bibitem{zhao2024scenario}
	Y.~Zhao, J.~T. Du, J.~Chen, Scenario-based adaptations of differential privacy:
	A technical survey, ACM Computing Surveys (ACM Comput. Surv.) 56~(8) (2024)
	199:1--199:39.
	
	\bibitem{DBLP:conf/ccs/ErlingssonPK14}
	{\'{U}}.~Erlingsson, V.~Pihur, A.~Korolova,
	%\href{http://doi.acm.org/10.1145/2660267.2660348}
	{{RAPPOR:} randomized aggregatable privacy-preserving ordinal response}, 
	in: Proceedings of the 2014 {ACM} {SIGSAC} Conference on Computer and Communications Security, Scottsdale, AZ, USA, November 3-7, 2014, 2014, pp. 1054--1067.
	% \newblock %\href {https://doi.org/10.1145/2660267.2660348}
	% {\path{doi:10.1145/2660267.2660348}}.
	% \newline\urlprefix\url{http://doi.acm.org/10.1145/2660267.2660348}
	
	\bibitem{DBLP:journals/concurrency/0055ZL022}
	T.~Zhang, T.~Zhu, R.~Liu, W.~Zhou,
	%\href{https://doi.org/10.1002/cpe.6015}
	{Correlated data in differential privacy: Definition and analysis}, 
	Concurr. Comput. Pract. Exp. 34~(16) (2022).
	% \newblock %\href {https://doi.org/10.1002/CPE.6015}
	% {\path{doi:10.1002/CPE.6015}}.
	% \newline\urlprefix\url{https://doi.org/10.1002/cpe.6015}
	
	\bibitem{DBLP:conf/crypto/GehrkeHLP12}
	J.~Gehrke, M.~Hay, E.~Lui, R.~Pass,
	%\href{http://dx.doi.org/10.1007/978-3-642-32009-5_28}
	{Crowd-blending privacy}, 
	in: Advances in Cryptology - {CRYPTO} 2012 - 32nd Annual Cryptology Conference, Santa Barbara, CA, USA, August 19-23, 2012. Proceedings, 2012, pp. 479--496.
	% \newblock %\href {https://doi.org/10.1007/978-3-642-32009-5_28}
	% {\path{doi:10.1007/978-3-642-32009-5_28}}.
	% \newline\urlprefix\url{http://dx.doi.org/10.1007/978-3-642-32009-5_28}
	
	\bibitem{DBLP:journals/ijufks/Sweene02}
	L.~Sweeney, %\href{http://dx.doi.org/10.1142/S0218488502001648}
	{k-anonymity: {A} model for protecting privacy}, 
	International Journal of Uncertainty, Fuzziness and Knowledge-Based Systems 10~(5) (2002) 557--570.
	% \newblock %\href {https://doi.org/10.1142/S0218488502001648}
	% {\path{doi:10.1142/S0218488502001648}}.
	% \newline\urlprefix\url{http://dx.doi.org/10.1142/S0218488502001648}
	
	\bibitem{DBLP:journals/tods/KiferM14}
	D.~Kifer, A.~Machanavajjhala,
	%\href{http://doi.acm.org/10.1145/2514689}
	{Pufferfish: {A} framework for mathematical privacy definitions}, 
	{ACM} Trans. Database Syst. 39~(1) (2014) 3.
	% \newblock %\href {https://doi.org/10.1145/2514689} {\path{doi:10.1145/2514689}}.
	% \newline\urlprefix\url{http://doi.acm.org/10.1145/2514689}
	
	\bibitem{DBLP:conf/focs/BassilyGKS13}
	R.~Bassily, A.~Groce, J.~Katz, A.~D. Smith,
	%\href{https://doi.org/10.1109/FOCS.2013.54}
	{Coupled-worlds privacy: Exploiting adversarial uncertainty in statistical data privacy}, 
	in: 54th Annual {IEEE} Symposium on Foundations of Computer Science, {FOCS} 2013, 26-29 October, 2013, Berkeley, CA, {USA}, 2013, pp. 439--448.
	% \newblock %\href {https://doi.org/10.1109/FOCS.2013.54}
	% {\path{doi:10.1109/FOCS.2013.54}}.
	% \newline\urlprefix\url{https://doi.org/10.1109/FOCS.2013.54}
	
	\bibitem{DBLP:conf/tcc/GehrkeLP11}
	J.~Gehrke, E.~Lui, R.~Pass,
	%\href{http://dx.doi.org/10.1007/978-3-642-19571-6_26}
	{Towards privacy for social networks: {A} zero-knowledge based definition of privacy}, 
	in: Theory of Cryptography - 8th Theory of Cryptography Conference, {TCC} 2011, Providence, RI, USA, March 28-30, 2011. Proceedings, 2011, pp. 432--449.
	% \newblock %\href {https://doi.org/10.1007/978-3-642-19571-6_26}
	% {\path{doi:10.1007/978-3-642-19571-6_26}}.
	% \newline\urlprefix\url{http://dx.doi.org/10.1007/978-3-642-19571-6_26}
	
	\bibitem{DBLP:journals/corr/DworkR16}
	C.~Dwork, G.~N. Rothblum, %\href{http://arxiv.org/abs/1603.01887}
	{Concentrated differential privacy}, 
	CoRR abs/1603.01887 (2016).
	% \newline\urlprefix\url{http://arxiv.org/abs/1603.01887}
	
	\bibitem{DBLP:conf/tcc/BunS16}
	M.~Bun, T.~Steinke,
	%\href{https://doi.org/10.1007/978-3-662-53641-4_24}
	{Concentrated differential privacy: Simplifications, extensions, and lower bounds}, 
	in: Theory of Cryptography - 14th International Conference, {TCC} 2016-B, Beijing, China, October 31 - November 3, 2016, Proceedings, Part {I}, 2016, pp. 635--658.
	% \newblock %\href {https://doi.org/10.1007/978-3-662-53641-4_24}
	% {\path{doi:10.1007/978-3-662-53641-4_24}}.
	% \newline\urlprefix\url{https://doi.org/10.1007/978-3-662-53641-4_24}
	
	\bibitem{DBLP:journals/popets/DesfontainesP20}
	D.~Desfontaines, B.~Pej{\'{o}},
	%\href{https://doi.org/10.2478/popets-2020-0028}
	{Sok: Differential privacies},
	Proc. Priv. Enhancing Technol. 2020~(2) (2020) 288--313.
	% \newblock %\href {https://doi.org/10.2478/POPETS-2020-0028}
	% {\path{doi:10.2478/POPETS-2020-0028}}.
	% \newline\urlprefix\url{https://doi.org/10.2478/popets-2020-0028}
	
	\bibitem{dalenius1977towards}
	T.~Dalenius,
	%\href{https://www.vrdc.cornell.edu/info7470/2011/Readings/dalenius-1977.pdf}
	{{Towards a methodology for statistical disclosure control}}, 
	Statistik Tidskrift 5~(2--1) (1977) 429--444.
	% \newline\urlprefix\url{https://www.vrdc.cornell.edu/info7470/2011/Readings/dalenius-1977.pdf}
	
	\bibitem{DBLP:conf/icalp/Dwork06}
	C.~Dwork, Differential privacy, in: ICALP (2), 2006, pp. 1--12.
	
	\bibitem{6769090}
	C.~E. Shannon, Communication theory of secrecy systems, The Bell System
	Technical Journal 28~(4) (1949) 656--715.
	% \newblock %\href {https://doi.org/10.1002/j.1538-7305.1949.tb00928.x}
	% {\path{doi:10.1002/j.1538-7305.1949.tb00928.x}}.
	
	\bibitem{DBLP:books/daglib/0016881}
	T.~M. Cover, J.~A. Thomas,
	%\href{http://www.elementsofinformationtheory.com/}
	{Elements of information theory {(2.} ed.)}, 
	Wiley, 2006.
	% \newline\urlprefix\url{http://www.elementsofinformationtheory.com/}
	
	\bibitem{DBLP:journals/jcss/GoldwasserM84}
	S.~Goldwasser, S.~Micali,
	%\href{https://doi.org/10.1016/0022-0000(84)90070-9}
	{Probabilistic encryption}, 
	J. Comput. Syst. Sci. 28~(2) (1984) 270--299.
	% \newblock %\href {https://doi.org/10.1016/0022-0000(84)90070-9}
	% {\path{doi:10.1016/0022-0000(84)90070-9}}.
	% \newline\urlprefix\url{https://doi.org/10.1016/0022-0000(84)90070-9}
	
	\bibitem{DBLP:conf/focs/Yao82a}
	A.~C. Yao, %\href{https://doi.org/10.1109/SFCS.1982.45}
	{Theory and applications of trapdoor functions (extended abstract)}, 
	in: 23rd Annual Symposium on Foundations of Computer Science, Chicago, Illinois, USA, 3-5 November 1982, 1982, pp. 80--91.
	% \newblock %\href {https://doi.org/10.1109/SFCS.1982.45}
	% {\path{doi:10.1109/SFCS.1982.45}}.
	% \newline\urlprefix\url{https://doi.org/10.1109/SFCS.1982.45}
	
	\bibitem{DBLP:conf/focs/RogersRST16}
	R.~M. Rogers, A.~Roth, A.~D. Smith, O.~Thakkar,
	%\href{https://doi.org/10.1109/FOCS.2016.59}
	{Max-information, differential privacy, and post-selection hypothesis testing}, 
	in: {IEEE} 57th Annual Symposium on Foundations of Computer Science, {FOCS} 2016, 9-11 October 2016, Hyatt Regency, New Brunswick, New Jersey, {USA}, 2016, pp. 487--494.
	% \newblock %\href {https://doi.org/10.1109/FOCS.2016.59}
	% {\path{doi:10.1109/FOCS.2016.59}}.
	% \newline\urlprefix\url{https://doi.org/10.1109/FOCS.2016.59}
	
	\bibitem{DBLP:conf/tcc/BunCV16}
	M.~Bun, Y.~Chen, S.~P. Vadhan,
	%\href{https://doi.org/10.1007/978-3-662-53641-4_23}
	{Separating computational and statistical differential privacy in the client-server model}, 
	in: Theory of Cryptography - 14th International Conference, {TCC} 2016-B, Beijing, China, October 31 - November 3, 2016, Proceedings, Part {I}, 2016, pp. 607--634.
	% \newblock %\href {https://doi.org/10.1007/978-3-662-53641-4_23}
	% {\path{doi:10.1007/978-3-662-53641-4_23}}.
	% \newline\urlprefix\url{https://doi.org/10.1007/978-3-662-53641-4_23}
	
	\bibitem{DBLP:journals/tit/Blahut72}
	R.~E. Blahut, %\href{https://doi.org/10.1109/TIT.1972.1054855}
	{Computation of channel capacity and rate-distortion functions}, 
	{IEEE} Trans. Inf. Theory 18~(4) (1972) 460--473.
	% \newblock %\href {https://doi.org/10.1109/TIT.1972.1054855}
	% {\path{doi:10.1109/TIT.1972.1054855}}.
	% \newline\urlprefix\url{https://doi.org/10.1109/TIT.1972.1054855}
	
	\bibitem{li2023private}
	Y.~Li, M.~Purcell, T.~Rakotoarivelo, D.~Smith, T.~Ranbaduge, K.~S. Ng, Private
	graph data release: A survey, ACM Computing Surveys (ACM Comput. Surv.)
	55~(11) (2023) 226:1--226:39.
	% \newblock %\href {https://doi.org/10.1145/3573384} {\path{doi:10.1145/3573384}}.
	
	\bibitem{DBLP:conf/ccs/LiQSWY13}
	N.~Li, W.~H. Qardaji, D.~Su, Y.~Wu, W.~Yang,
	%\href{http://doi.acm.org/10.1145/2508859.2516686}
	{Membership privacy: a unifying framework for privacy definitions}, 
	in: 2013 {ACM} {SIGSAC} Conference on Computer and Communications Security, CCS'13, Berlin, Germany, November 4-8, 2013, 2013, pp. 889--900.
	% \newblock %\href {https://doi.org/10.1145/2508859.2516686}
	% {\path{doi:10.1145/2508859.2516686}}.
	% \newline\urlprefix\url{http://doi.acm.org/10.1145/2508859.2516686}
	
	\bibitem{zheng2025benchmarking}
	Z.~Z, Y.~L, H.~H, W.~G, Benchmarking relaxed differential privacy in private
	learning: A comparative survey, ACM Computing Surveys 57~(12) (2025) 1--34.
	% \newblock %\href {https://doi.org/10.1145/3729216} {\path{doi:10.1145/3729216}}.
	
	\bibitem{DBLP:conf/icde/MachanavajjhalaGKV06}
	A.~Machanavajjhala, J.~Gehrke, D.~Kifer, M.~Venkitasubramaniam,
	%\href{http://dx.doi.org/10.1109/ICDE.2006.1}
	{l-diversity: Privacy beyond k-anonymity}, 
	in: Proceedings of the 22nd International Conference on Data Engineering, {ICDE} 2006, 3-8 April 2006, Atlanta, GA, {USA}, 2006, p.~24.
	% \newblock %\href {https://doi.org/10.1109/ICDE.2006.1}
	% {\path{doi:10.1109/ICDE.2006.1}}.
	% \newline\urlprefix\url{http://dx.doi.org/10.1109/ICDE.2006.1}
	
	\bibitem{DBLP:conf/icde/LiLV07}
	N.~Li, T.~Li, S.~Venkatasubramanian,
	%\href{http://dx.doi.org/10.1109/ICDE.2007.367856}
	{t-closeness: Privacy beyond k-anonymity and l-diversity}, 
	in: Proceedings of the 23rd International Conference on Data Engineering, {ICDE} 2007, The Marmara Hotel, Istanbul, Turkey, April 15-20, 2007, 2007, pp. 106--115.
	% \newblock %\href {https://doi.org/10.1109/ICDE.2007.367856}
	% {\path{doi:10.1109/ICDE.2007.367856}}.
	% \newline\urlprefix\url{http://dx.doi.org/10.1109/ICDE.2007.367856}
	
	\bibitem{whitehouse2023fully}
	J.~Whitehouse, A.~Ramdas, R.~Rogers, S.~Wu,
	%\href{https://arxiv.org/abs/2203.05481}
	{Fully-adaptive composition in differential privacy}, 
	in: The 40th International Conference on Machine Learning (ICML 2023), 2023.
	% \newblock %\href {https://doi.org/10.48550/arXiv.2203.05481}
	% {\path{doi:10.48550/arXiv.2203.05481}}.
	% \newline\urlprefix\url{https://arxiv.org/abs/2203.05481}
	
	\bibitem{DBLP:conf/forc/LigettPR20}
	K.~Ligett, C.~Peale, O.~Reingold,
	%\href{https://doi.org/10.4230/LIPIcs.FORC.2020.10}
	{Bounded-leakage differential privacy}, 
	in: A.~Roth (Ed.), 1st Symposium on Foundations of Responsible Computing, {FORC} 2020, June 1-3, 2020, Harvard University, Cambridge, MA, {USA} (virtual conference), Vol. 156 of LIPIcs, Schloss Dagstuhl - Leibniz-Zentrum f{\"{u}}r Informatik, 2020, pp. 10:1--10:20.
	% \newblock %\href {https://doi.org/10.4230/LIPIcs.FORC.2020.10}
	% {\path{doi:10.4230/LIPIcs.FORC.2020.10}}.
	% \newline\urlprefix\url{https://doi.org/10.4230/LIPIcs.FORC.2020.10}
	
	\bibitem{DBLP:conf/ndss/LiuMC16}
	C.~Liu, P.~Mittal, S.~Chakraborty,
	%\href{http://www.internetsociety.org/sites/default/files/blogs-media/dependence-makes-you-vulnerable-differential-privacy-under-dependent-tuples.pdf}
	{Dependence makes you vulnerable: Differential privacy under dependent tuples}, 
	in: 23nd Annual Network and Distributed System Security Symposium, {NDSS} 2016, San Diego, California, USA, February 21-24, 2016, 2016.
	% \newline\urlprefix\url{http://www.internetsociety.org/sites/default/files/blogs-media/dependence-makes-you-vulnerable-differential-privacy-under-dependent-tuples.pdf}
	
	\bibitem{DBLP:conf/ccs/CuffY16}
	P.~Cuff, L.~Yu, %\href{http://doi.acm.org/10.1145/2976749.2978308}
	{Differential privacy as a mutual information constraint}, 
	in: Proceedings of the 2016 {ACM} {SIGSAC} Conference on Computer and Communications Security, Vienna, Austria, October 24-28, 2016, 2016, pp. 43--54.
	% \newblock %\href {https://doi.org/10.1145/2976749.2978308}
	% {\path{doi:10.1145/2976749.2978308}}.
	% \newline\urlprefix\url{http://doi.acm.org/10.1145/2976749.2978308}
	
	\bibitem{DBLP:journals/tit/WangYZ16}
	W.~Wang, L.~Ying, J.~Zhang,
	%\href{http://dx.doi.org/10.1109/TIT.2016.2584610}
	{On the relation between identifiability, differential privacy, and mutual-information privacy},
	{IEEE} Trans. Information Theory 62~(9) (2016) 5018--5029.
	% \newblock %\href {https://doi.org/10.1109/TIT.2016.2584610}
	% {\path{doi:10.1109/TIT.2016.2584610}}.
	% \newline\urlprefix\url{http://dx.doi.org/10.1109/TIT.2016.2584610}
	
	\bibitem{wu2021achievingdalenius}
	G.~Wu, X.~Xia, Y.~He, %\href{https://arxiv.org/abs/1703.07474}
	{Achieving dalenius' goal of data privacy with practical assumptions} (2021).
	 \newblock \href {http://arxiv.org/abs/1703.07474} {\path{arXiv:1703.07474}},
	% \href {https://doi.org/10.48550/arXiv.1703.07474}
	% {\path{doi:10.48550/arXiv.1703.07474}}.
	% \newline\urlprefix\url{https://arxiv.org/abs/1703.07474}
	
\end{thebibliography}



\end{document}